\documentclass[prl,aps,groupedaddress,onecolumn,floatfix,showpacs]{revtex4-1}

\pdfoutput=1
\usepackage{graphicx}
\usepackage{dcolumn}
\usepackage{bm}
\usepackage{times}
\usepackage{amsmath}
\usepackage{epsfig}
\usepackage{dcolumn}
\usepackage{amssymb}
\usepackage{color}

\begin{document}


\title{Unexpectedly high pressure for molecular dissociation in liquid hydrogen by a reliable electronic simulation} 

\author{Guglielmo Mazzola$^{1,2}$}
\email{gmazzola@sissa.it}
\author{Seiji Yunoki$^{3,4,5}$}
\author{Sandro Sorella$^{1,2,3}$}
\affiliation{
$^1$SISSA -- International School for Advanced Studies, Via Bonomea 265, 34136 Trieste, Italy\\
$^2$Democritos Simulation Center CNR--IOM Istituto Officina dei Materiali, Via Bonomea 265, 34136 Trieste, Italy\\
$^3$Computational Materials Science Research Team, RIKEN Advanced Institute for 
Computational Science (AICS), Kobe, Hyogo 650-0047, Japan\\
$^4$Computational Condensed Matter Physics Laboratory, RIKEN, Wako, Saitama 351-0198, Japan\\
$^5$Computational Quantum Matter Research Team, RIKEN Center for Emergent Matter Science (CEMS), Wako, Saitama 351-0198, Japan}

\begin{abstract}
  The study of the high pressure phase diagram of hydrogen has continued with renewed effort for about one century as it remains a fundamental challenge for experimental and theoretical techniques. Here we employ an efficient molecular dynamics based on the quantum Monte Carlo method, which can describe accurately the electronic correlation and treat a large number of hydrogen atoms, allowing a realistic and reliable prediction of thermodynamic properties.
We find that the molecular liquid phase is unexpectedly stable and the transition towards  a fully atomic liquid phase occurs at much higher pressure than previously believed. The old standing problem of low temperature atomization is, therefore, still far from experimental reach. 
\end{abstract}

\maketitle

\begin{figure}[h]
\begin{center}
\includegraphics[scale=0.53]{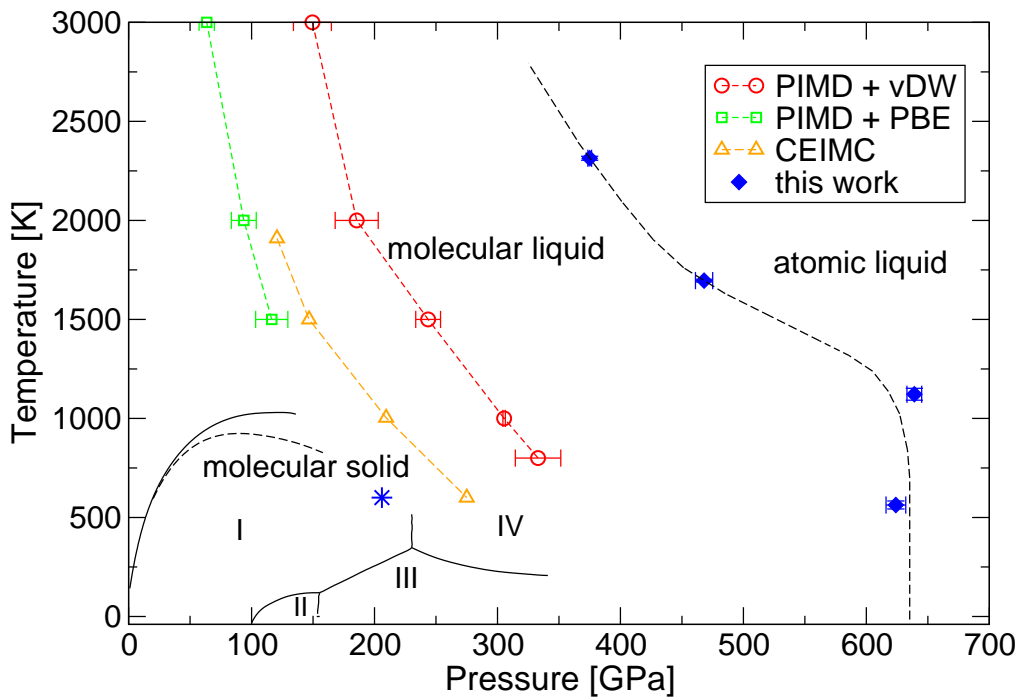}
\caption{ {\bf $P$-$T$ phase diagram of hydrogen.} Black solid lines indicate experimental boundaries between the molecular liquid and the molecular solid, the latter consisting of four different solid phases denoted by I, II, III, and IV as in Ref.~\cite{hemley}. Colored symbols with dashed curves correspond to the liquid-liquid transition (LLT) obtained with latest simulations. Red circles and green squares refer to Density Functional Theory (DFT) calculations with different functionals (PBE and vdW)~\cite{mcmahon} while orange triangles refer to Coupled Electron-Ion Monte Carlo (CEIMC)~\cite{morales,lib}. Blue diamonds correspond to the LLT estimated in this work. Our simulations also find that solidification occurs starting from a molecular liquid at a parameter indicated by a blue star. The black dashed line is a guide to the eye. 
}
\label{fig:phase}
\end{center}
\end{figure}

Hydrogen is the simplest element in nature and nevertheless its phase diagram at high pressures ($P$s) remains a challenge both from the experimental and theoretical points of view. Moreover, the understanding of equilibrium properties of hydrogen in the high pressure regime is crucial for a satisfactory description of many astrophysical bodies~\cite{astro} and for discovering new phases  in condensed matter systems. At normal pressures and temperatures ($T$s) the hydrogen molecule H$_2$ is exceptionally stable and thus the usual phases are described in terms of these molecules, i.e., solid, liquid, and gas molecular phases (see Fig.\ref{fig:phase}). 

In the early days, it was conjectured by Wigner and Huntington~\cite{wigner} that, upon high pressure,  this stable entity -- the H$_2$ molecule -- can be destabilized, giving rise to an electronic system composed of one electron for each localized atomic center, namely, the condition that, according to the band theory, should lead to metallic behavior. After this conjecture, an extensive experimental and theoretical effort has been devoted, an effort that continues to be particularly active also recently~\cite{revhyd, weir,eremets,nellis,fortov2007,dzy}, where some evidence of metallicity in hydrogen has been reported. While a small resistivity has been observed in the molecular liquid state in the range of $140 - 180$ GPa and $2000 - 3000$ K~\cite{weir}, it is not clear whether this observation is above a critical temperature $T^*$, where only a smooth metal-insulator crossover can occur, and whether metallization is due to dissociation or can occur even within the molecular phase. Evidence of a phase transition has been clearly reported in Ref.~\cite{fortov2007}, though the temperature has not been measured directly. On the other hand~\cite{dzy},  indirect evidence of a phase transition  at around 120 GPa and 1500 K has been claimed. By contrast, any indication of metallization has not been observed in the low-temperature solid phase yet~\cite{goncha,occelli,hemley}.

Until very recently, the Density Functional Theory (DFT) method has been considered the standard tool for the simulation of electronic phases, because it allows the simulation of many electrons with a reasonable computational effort. However, there are several drawbacks in this technique especially for the study of the dissociation of hydrogen: i) the single molecule is not accurately described at equilibrium and especially in the dissociation limit~\cite{dfth2,rewchemdft} (see Fig.\ref{fig:finitesize}c). ii) Electronic gaps are substantially underestimated~\cite{revdft} within DFT, implying that possible molecular phases are more easily destabilized within standard DFs. For all the above reasons, DFT seems not adequate for the hydrogen problem under high pressure, especially in a range of pressures unaccessible by experiments, where the quality of a particular DF cannot be validated. Indeed, several DFT simulations on this particular subject~\cite{scandolo, bonev1, tamblyn, holts} lead to contradictory results for the nature of the molecular liquid-atomic liquid transition and its position in the phase diagram may vary in a range of more than 100 GPa according to different DFs~\cite{bonev2, mcmahon, morales}. Recently, it has also been shown that DFT solid stable phases  strongly depend on the DF used~\cite{azadi,azadi2,pierpa}, suggesting quite clearly that the predictive power of DFT is limited for hydrogen. 
 
Among all first principles simulation methods, the quantum Monte Carlo (QMC) method provides a good balance between accuracy and computational cost and it appears very suitable for this problem. The QMC approach is based on a many-body wave function -- the so called trial wave function -- and no approximation is used in the exchange and correlation contributions within the given ansatz~\cite{qmc}.  For instance, the exact equilibrium bond length and the dissociation energy profile for the isolated $H_2$ molecule are correctly recovered (see Fig.\ref{fig:finitesize}.c) within this approach. Only few years ago the calculation of forces has been made possible within QMC, with affordable algorithms~\cite{attaccalite}. Therefore, very efficient sampling methods are now possible, that are based on molecular dynamics (MD), where in a single step all atomic coordinates are changed according to high quality forces corresponding to a correlated Born-Oppenheimer energy surface (see Methods). In this way, it is possible to employ long enough simulations that are well equilibrated and are independent of the initialization, also for very large size. 

With this newly developed QMC simulations, here we report numerical evidences for a first order liquid-liquid transition (LLT) between the molecular and the atomic fluids. We find critical behavior in pressures as a function of (isothermal) density and as a function of (isochoric) temperature. Moreover, a clear abrupt change of the radial pair distribution function at the LLT is also observed, even though molecules start to gradually dissociate well before the transition. We also find that the LLT occurs at much higher densities, hence at much higher pressures, as compared with recent calculations~\cite{morales, mcmahon}. By looking at the steepness of the evaluated LLT boundary in the phase diagram (see Fig.~\ref{fig:phase}), we can also give the estimate of $\sim 600$ GPa for the expected complete atomization of the zero-temperature structure. However this remains an open issue since in our work we have neglected quantum effects on protons, that may be certainly important at low temperatures. Moreover this quantitative prediction should hold provided the dissociation occurs between two liquid phases down to zero temperature\cite{labet} and no solid atomic phase emerges.

\subsection{Results}

{\bf System size and sampling scheme.} We employ simulation cells containing up to $256$ hydrogen atoms and use a novel MD scheme with friction in the NVT ensemble (see Supplementary Note 1, Supplementary Fig. 1), for simulation times of few picoseconds, long enough to have well converged results on the pressure, internal energy, and the radial pair distribution function $g(r)$. 

\begin{figure}[h!]
\begin{center}
\includegraphics[scale=0.40]{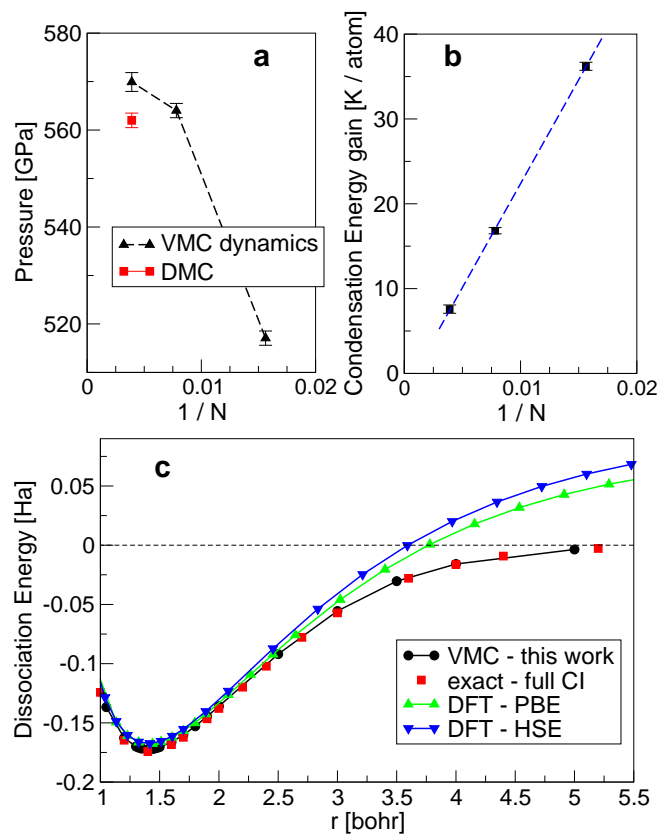}
\caption{ {\bf Accuracy and finite size effects.} ({\bf a}) Pressure as a function of the system sizes $N = $64, 128, and 256 at a density 
$r_s=1.22$ (the Wigner-Seitz radius $r_s$ is defined as $V/N= 4/3 \pi (r_s a_0)^3$where $V$ is the volume, $N$ the number of ions, and $a_0$
is the Bohr radius.) near the transition at 600 K. The Diffusion Monte Carlo (DMC) value, obtained from  
20  equilibrated ($N=256$) configurations generated by the Variational Monte Carlo (VMC) dynamics,  is also plotted (red square). ({\bf b}) Finite size scaling of the condensation energy gain at $r_s= 1.28$ and 600 K. The condensation energy gain becomes negligible in the infinite size limit (see Methods). ({\bf c}) Dissociation energy curves for the $H_2$ molecule for different methods, QMC at the VMC level and with the same wavefunction variational ansatz employed in the dynamics, DFT with PBE or HSE DFs\cite{revdft}, and the exact curve obtained with full configuration interaction (CI) method\cite{corongiu}.}
\label{fig:finitesize}
\end{center}
\end{figure}

Notice also that our approximation to consider QMC in its simplest variational formulation (see Methods), i.e., variational Monte Carlo (VMC), is already quite satisfactory because the much more computationally expensive diffusion Monte Carlo (DMC) can correct the pressure only by few GPa's (see Fig.\ref{fig:finitesize}a) 
which is not relevant for the present accuracy in the phase diagram, rather size effects seem to be much more important (see Methods). 

In this regard we also performed a DMC-MD on a much smaller system with 54 protons and we  have verified that, apart from an overall shift in the total electronic energy, VMC and DMC dynamics give quantitatively the same results for the pressure and the $g(r)$ (see Supplementary Fig. 2, Supplementary Note 2). This implies that the forces, evaluated at the VMC level, are already accurate to drive the dynamics to the correct equilibrium distribution.

\begin{figure}[h!]
\begin{center}

\includegraphics[scale=0.525]{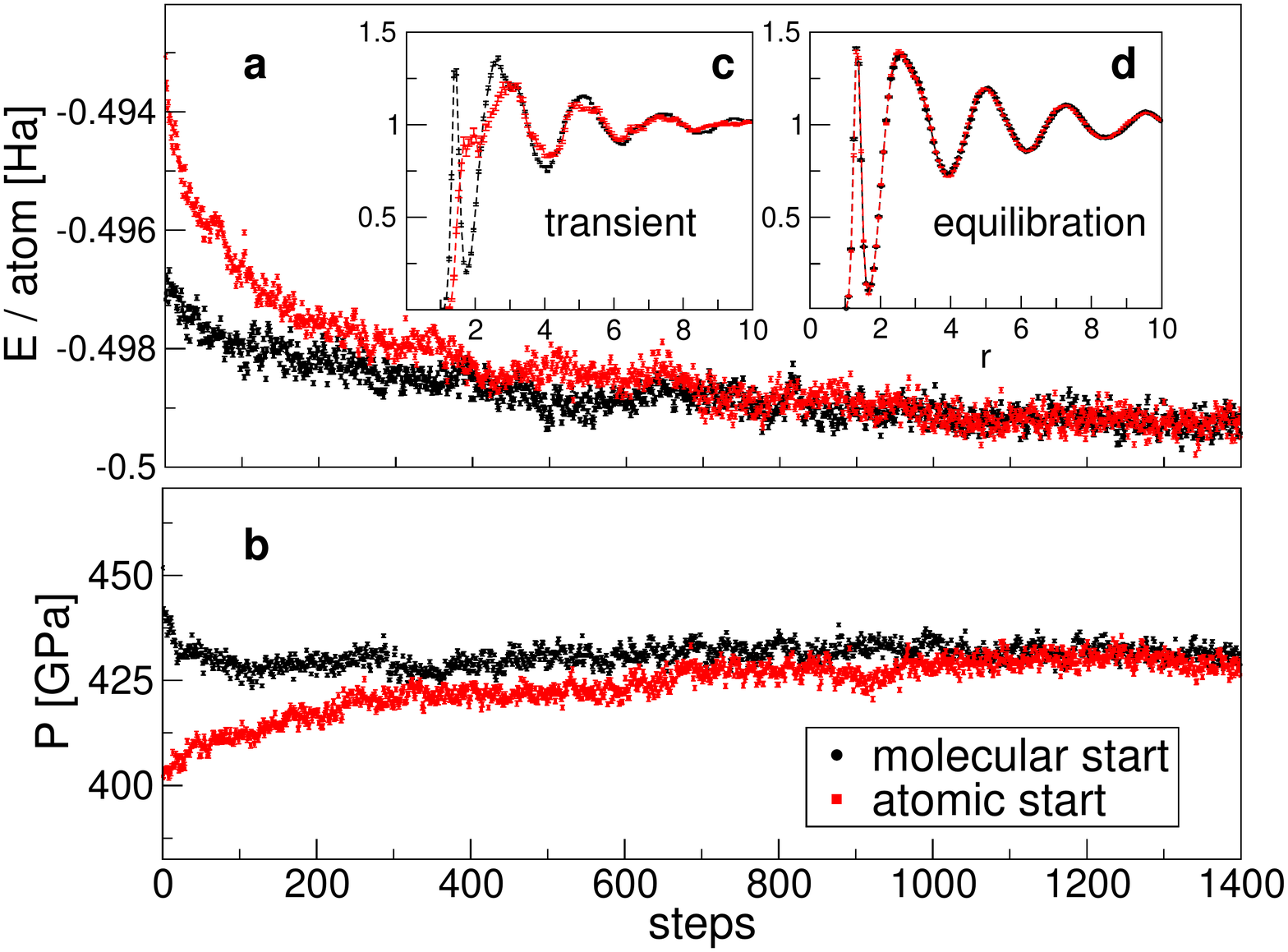}
\caption{{\bf Ergodicity of the simulations.} ({\bf a}) Energy $E$ and ({\bf b}) pressure $P$ as a function of the ionic steps 
during a Langevin dynamics simulation at $r_s=1.28$ and $T=600$ K. 
Red squares refer to a simulation whose starting configuration is an atomic fluid, while black circles correspond to a molecular 
initial configuration (see inset {\bf c} for the radial pair distribution functions of these two configurations). After a short equilibration, 
energy, pressure, and radial pair distribution function (inset {\bf d}) converge to the same values. 
}
\label{fig:ergo}
\end{center}
\end{figure}

{\bf Ergodicity of QMC simulations.} In our approach, we cannot address directly the issue of metallicity or insulator behavior but can assess possible first order transition by large scale simulations, where any discontinuity of correlation functions with varying temperature or pressure should be fairly evident and clear. Considering that even an insulator with a true gap should also exhibit residual activated conductivity at finite temperatures, a metal-insulator transition at finite temperatures can be experimentally characterized only by a jump (usually by several orders of magnitude) of the conductivity. Hence, without a first order transition, namely, without discontinuous jumps in the correlation functions upon a smooth variation of temperature or pressure, we cannot define a true metal-insulator transition at finite temperatures, otherwise it is rather a crossover.

In order to assess that our QMC method is capable of characterizing correctly a phase transition, we carefully check for the possible lack of thermalization near the phase transition by repeating the simulations with very different starting ionic configurations at the same thermodynamic point. 
Moreover, the functional form of the trial wave function is flexible enough to correctly describe both the paired and the dissociated state (see Methods), and therefore our approach is expected to be particularly accurate even for the LLT. A complete equilibration is reached within the QMC framework since no hysteresis effects occur in all the range of temperatures studied (see Fig.~\ref{fig:ergo} and Supplementary Fig. 3).

\begin{figure}[h!]
\begin{center}
\includegraphics[scale=0.40]{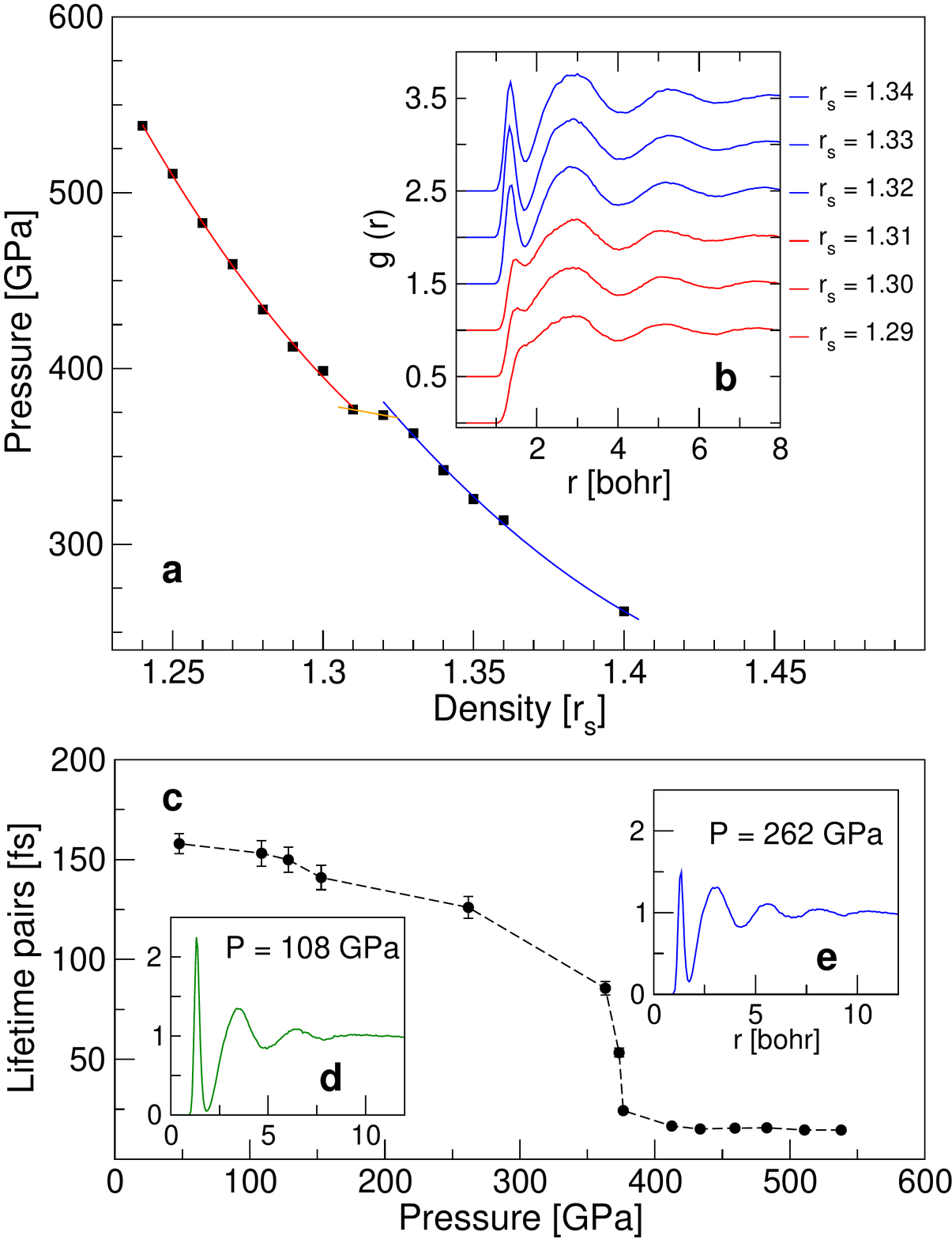}
\caption{{\bf Liquid-liquid transition at 2300 K.} ({\bf a}) Pressure as a function of the density. A clear plateau is visible around $r_s=1.31 - 1.32$, 
indicating the first order transition. This evidence is also supported by the discontinuous change with $r_s$ in the radial pair distribution function $g(r)$ (inset {\bf b}). ({\bf c}) Average lifetime of pairs as a function of pressure. A pair is defined here as a couple of ions which are nearest neighbors and whose distance $r$ is smaller than a cutoff $r_c= 1.70$ a.u. The shape of this curve is qualitatively similar for every reasonable choice of $r_c$ although its amplitude may slightly vary. In the insets ({\bf d}) and ({\bf e}), radial pair distribution functions for two different pressures ($P$) in the molecular fluid. The higher the pressure, the smaller is the molecular peak and more coordination shells appear in the long range tail. }
\label{fig:2400}
\end{center}
\end{figure}

{\bf Characterizing the first-order transition. }
To identify the LLT, we trace four isotherms in the range 600 - 2300 K, looking for a possible singular behaviour of the pressure and the radial pair distribution function $g(r)$, in a wide range of density (see Supplementary Note 3, Supplementary Fig. from 4 to 11 ). We indeed find a relatively small discontinuity, which appears to be clear also at the highest temperature considered (see Fig.~\ref{fig:2400}a and b). A similar first-order behavior is also found by looking at the pressure as a function of temperature at fixed density (see Supplementary Fig. 12), i.e. by crossing the LLT vertically, along the isochor having density $r_s$=1.28. Notice that, close to the transition a fully molecular phase is not stable, as a large fraction of pairs is found to be already dissociated (Fig.~\ref{fig:2400}b). Our results are summarized in the $P$-$T$ phase diagram shown in Fig.~\ref{fig:phase}. We should note here that, in our calculations, we neglect nuclear quantum effects. However, this approximation should slightly  affect the location of the transition, as it was shown that the inclusion of the zero point motion shifts the LLT toward smaller pressures only by about 40 GPa~\cite{morales,mcmahon}. 

In the high pressure phase diagram, our results suggest the existence of a mixed -- although mainly molecular -- liquid, surrounding the solid IV mixed molecular atomic phase (see Fig.~\ref{fig:phase}). We have studied the average lifetime of the pairs formed by nearest neighboring hydrogen ions. As shown in Fig.~\ref{fig:2400}c, it exhibits a clear jump with varying pressure at 2300K, supporting the location of the LLT at 375 GPa. We also notice a precursor drop of the lifetime at around 150 GPa, which corresponds to the onset of the dissociation. This value of 150 GPa  is consistent with the pressure range where a drastic but continuous drop of the resistivity is observed   in the molecular phase\cite{weir}. In order to better characterize the LLT, we study in Fig.~\ref{fig:2400}d and e the dissociation fraction and the long range behavior of $g(r)$ for two fluid configurations at pressures much smaller than the true first order transition point. Nevertheless, a qualitative change in the behavior of these quantities is evident even within the same phase. Remarkably, not only the dissociation fraction increases with the pressure but also the number of oscillations in the long range $g(r)$ tail becomes larger, both features being very similar to what is observed in the high pressure phase. By taking also into account that, at non zero temperature, a finite and large conductivity can be activated, it is clear that a rather sharp variation of physical quantities can occur much before the first order transition.

\subsection{Discussion}

In conclusion, we have reported the first description of the dissociation transition in liquid hydrogen by {\it ab-initio} simulation based on QMC method with fairly large number of atoms. The main outcomes of our study are summarized as follows: i) the transition, which appears to be first-order, occurs at substantially higher pressures than the previous {\it ab-initio} predictions based on DFT. ii) Employing QMC simulations with large number of atoms is essential because the stability of the molecular phase is otherwise underestimated. iii) The first order character is evident also at the highest temperatures, suggesting that even at these temperatures 
this transition is not a crossover. iv) the shape of the LLT boundary is rather unusual and becomes a vertical line in the $P$-$T$ phase diagram for $T < 1100$ K. 
 By assuming that also at lower temperatures no solid phase emerges,  the dissociation pressure should remain almost temperature independent. Therefore, even by considering an upper limit of 100 GPa shift to lower pressures, due to proton quantum effects not included here, we predict that experiments should be done at  least above  500 Gpa to realize the Wigner and Huntington dream of hydrogen atomic metallization.

\section{Methods}
 
{\bf Accuracy of QMC methods.} 
In this work we employ the QMC approach for electronic properties. In the simplest formulation, the correlation between electrons is described by the so called Jastrow factor $J$ of the following general form
\begin{equation} 
\label{eq:jastrowsimple}
J = \prod \limits_{i<j}  e^{ u ({\bm r}_i,{\bm r}_j)  }
\end{equation} 
where $\bm r_i$ and $\bm r_j$ are electron positions and $u$ is a two-electron function to be determined variationally. It is enough to apply this factor to a single Slater determinant to remove the energetically expensive contributions of too close electrons occupying the same atomic orbital and to obtain for instance the correct dissociation limit for the $H_2$ molecule (see Fig. \ref{fig:finitesize}.c), and essentially exact results for the benchmark hydrogen chain model~\cite{stella}. Starting from this Jastrow correlated ansatz, an important projection scheme has been developed - the diffusion Monte Carlo (DMC) - that allows an almost exact description of the correlation energy, with a full {\it ab-initio} many-body approach, namely by the direct solution of the Schr\"odinger equation. Unfortunately, QMC is much more expensive than DFT, and so far its application has been limited to small number of atoms~\cite{morales}.

{\bf Variational wavefunction.}
For the calculation of the electronic energy and forces we use a trial correlated  wavefunction of the form \begin{equation}
 J |SD\rangle
\end{equation}
The determinantal part $|SD\rangle $ (Slater determinant) is constructed starting from $N/2$ molecular orbitals, $N$ being the total number of electrons, while the Jastrow part can be written as $J = J_1 J_2 J_3$. The Jastrow factor takes into account the electronic correlation between  electrons and is conventionally split
into a homogeneous interaction $J_2$ depending on the relative distance between two electrons i.e., a two-body term as in Eq.\ref{eq:jastrowsimple}, and a non homogeneous contribution depending also on the electron-ion distance, included in the one-body $J_1$ and three-body $J_3$. The exact functional form of these components is given in Ref.\cite{marchi}. Both the Jastrow functions and the determinant of molecular orbitals are expanded in a gaussian localized basis. The optimization of the molecular orbitals  is done simultaneously with the correlated Jastrow part. We performed a systematic reduction of the basis set in order to minimize the total number of variational parameters. Indeed for the present accuracy for the phase diagram a small $2s$/atom basis set is sufficient as a larger $3s1p$/atom basis set only improves the total energy of $<1$ mH/atom and leaves substantially unchanged the radial pair distribution function, i.e. the atomic(molecular) nature of the fluid (see Supplementary Fig. 13). The value for the LLT critic pressure is not significantly affected (for the present accuracy in the phase diagram) by the choice of the basis (see Supplementary Fig. 14), rather the difference between QMC and DFT with PBE functional is already evident for a system of 64 atoms. 

We now address an issue that, in our opinion, can be extremely important in the context of hydrogen metallization. Close to a metal-insulator transition a resonating valence bond scenario is possible, and may give rise to unconventional superconductivity\cite{pwa}, namely a superconductivity stabilized without the standard BCS electron-phonon mechanism. In order to study this interesting possibility, we have calculated the energy gain obtained by replacing the Slater determinant with a BCS type of wave function, both with the same form of the Jastrow factor. This quantity is known as the condensation energy~\cite{marchi}, and is non zero in the thermodynamic limit when the variational wave function represents a superconductor. Though we neglect quantum effects on protons and  we have not systematically studied this issue for all densities, this VMC condensation energy (see Fig.~\ref{fig:finitesize}b), computed by considering about 20 different independent samples at $r_s$=1.28 and T=600 K, is very small and decreases very quickly with $N$ by approaching zero in the thermodynamic limit ($1/N=0$). This result at least justifies the use of a simpler Slater determinant wave function, and shows that, the quality of the wave function can be hardly improved by different, in principle more accurate, variational ansatzes. In this way, a straightforward reduction of the number of variational parameters is possible,  by exploiting also  the fact that matrix elements connecting localized orbitals above a threshold distance $r_{\rm max}$ do not contribute significantly to the energy. Indeed, as we have systematically checked in several test cases (see Supplementary Fig. 15 and Supplementary Table 1), it is enough to consider $r_{\rm max}$= 4 a.u. to have essentially converged results for the molecular orbitals, implying a drastic reduction of the variational space (from $\simeq 40000$ parameters to $\simeq 5000$ in a 256 hydrogen system).

{\bf Finite size effects.}
All the results for the LLT presented here refer to a cubic supercell at the $\Gamma-$point (see Supplementary Table 2) with the largest affordable  number of atoms (256) in order to be as close as possible to the thermodynamic limit. Indeed, even tough the pressure seems to converge with the size of the simulation cell, the molecular (atomic) nature of the liquid is very sensitive to the number of atoms N. This issue was previously reported in Refs. \cite{bonev2,desj} and cannot be removed with a better \emph{k}-point sampling, because this will be equivalent to enforce a fictitious periodicity to a liquid phase. In particular, the N=64 supercell simulations, even with \emph{k}-point sampling, strongly favour the dissociated liquid in both DFT and QMC frameworks (see Supplementary Fig. 16). Thus the critical LLT density is severely underestimated by employing supercells smaller than N=256, which is now considered a standard size in DFT simulations of liquids. The main reason of this effect is the  structural frustration, requiring the use of much larger supercells, whose  dimension L has to exceed  the correlation length of the liquid. A possible rule of thumb consists in checking that the $g(r)$ is smoothly approaching its asymptotic value 1 at $r=L$. Our evidences support the conclusion that a failure in dealing with the finite size effects will result in a severe underestimation of the LLT critical densities. DFT-MD simulations were performed using the QuantumEspresso code\cite{qe}.

{\bf Sampling the canonical ensamble with Langevin dynamics.}
In this study, we sample the canonical equilibrium distribution for the ions by means of a second order generalized Langevin equation, as introduced in Ref.~\onlinecite{attaccalite}. The major advantage of this technique consists in the efficient control of the target temperature even in the presence of noisy QMC forces. Here we improve upon this scheme devising a numerical integrator which is affected by a smaller time step error (see Supplementary Note 1). We adopt the ground state Born-Oppenheimer approximation, namely the  variational parameters, defining our wave function, are all consistently  optimized at each iteration of our MD. Therefore the electronic entropy has been neglected in all our calculations. However we have carefully checked that this entropy contribution is clearly neglegible in the relevant temperature range studied (see Supplementary Note 4). 

As well known, hysteresis is usually found by using local updates in simple Monte Carlo schemes, that can not be therefore reliable to determine the phase boundary of a first order transition. Our method, based on an advanced second order MD with friction, is instead powerful enough that very different phases can be reached during the simulation, with time scales that remain accessible for feasible computations. Indeed, we have also experienced a spontaneous solidification in a pressure and temperature range where the solid phase is expected to be stable (see Fig.~\ref{fig:phase} and Supplementary Note 5, Supplementary Fig. 17 for details). 
Though this effect has been observed in a much smaller system (64 hydrogen atoms), we believe that, after the inclusion of proton quantum effects, the present method can also shed light on understanding low temperature solid phases, that remain still highly debated and controversial in recent years~\cite{pick}. 

\section{Acknowledgements}

We acknowledge  G. Bussi, S. Scandolo, F. Pederiva, S. Moroni, and S. De Gironcoli for useful discussions and support by MIUR, PRIN 2010-2011.
Computational resources are provided by PRACE Project Grant 2011050781 and K computer at RIKEN Advanced Institute for 
Computational Science (AICS). 

\section{Author contributions}

G.M. and S.S. designed the research 
and performed the QMC and DFT calculations. All authors conceived the project, partecipated in the discussion of the results and in the writing
 of the paper.


\newpage

\section{Supplementary Materials}

%



\begin{figure}[h!]
\begin{center}
\includegraphics[scale=0.50]{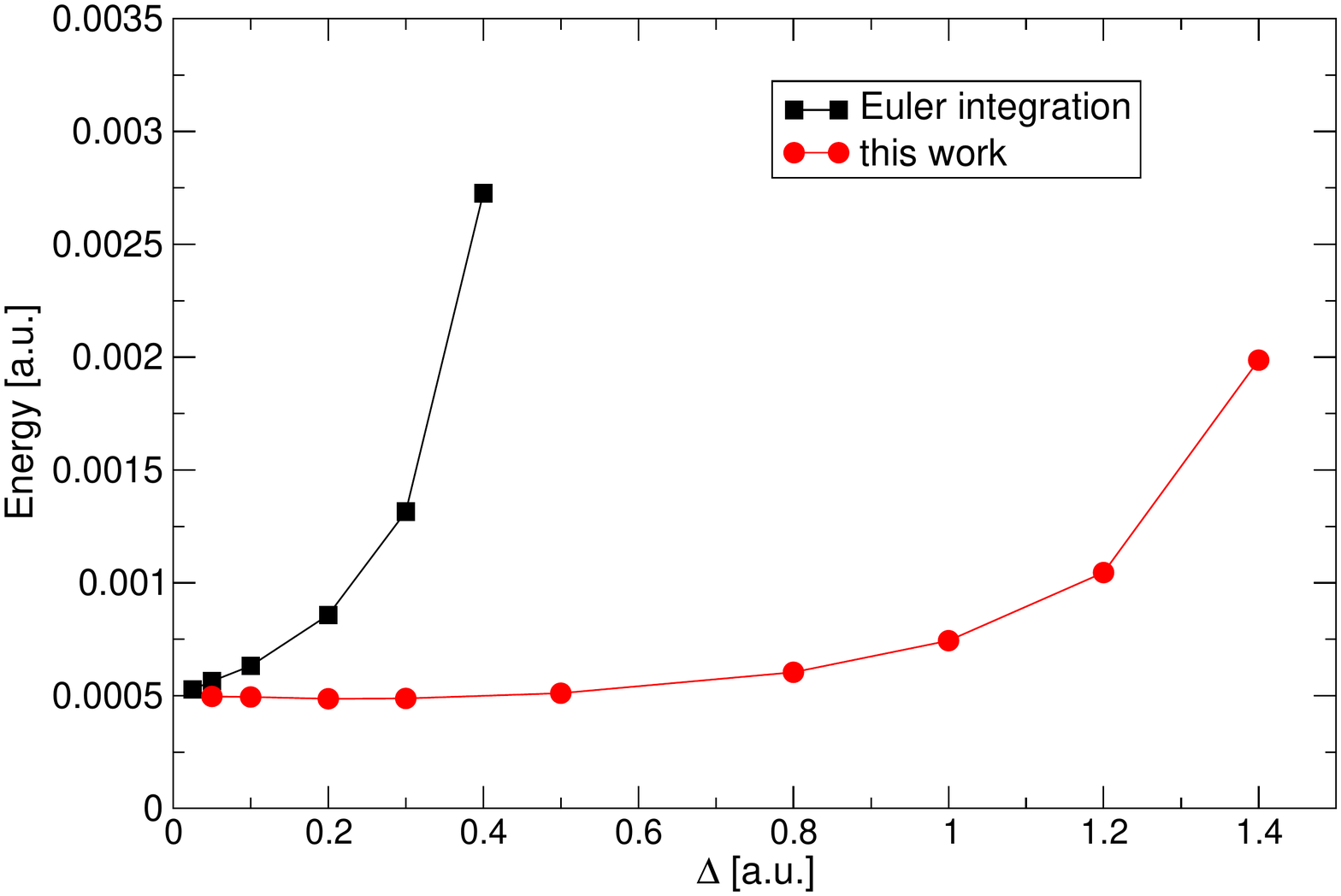}
\caption{{\bf Supplementary Figure} 1: Convergence of total energy as a function of the time step $\Delta$ for fixed friction and temperature (arbitrary unit). The toy model consists in a 2D particle subject to
a radial potential $U(r) = k (r - r_0)^2$. The new sampling scheme (red) is more accurate for larger $\Delta$ than the standard discretization scheme (black).}
\label{f:toy}
\end{center}
\end{figure}

\begin{figure}[h!]
\begin{center}
\includegraphics[scale=0.5]{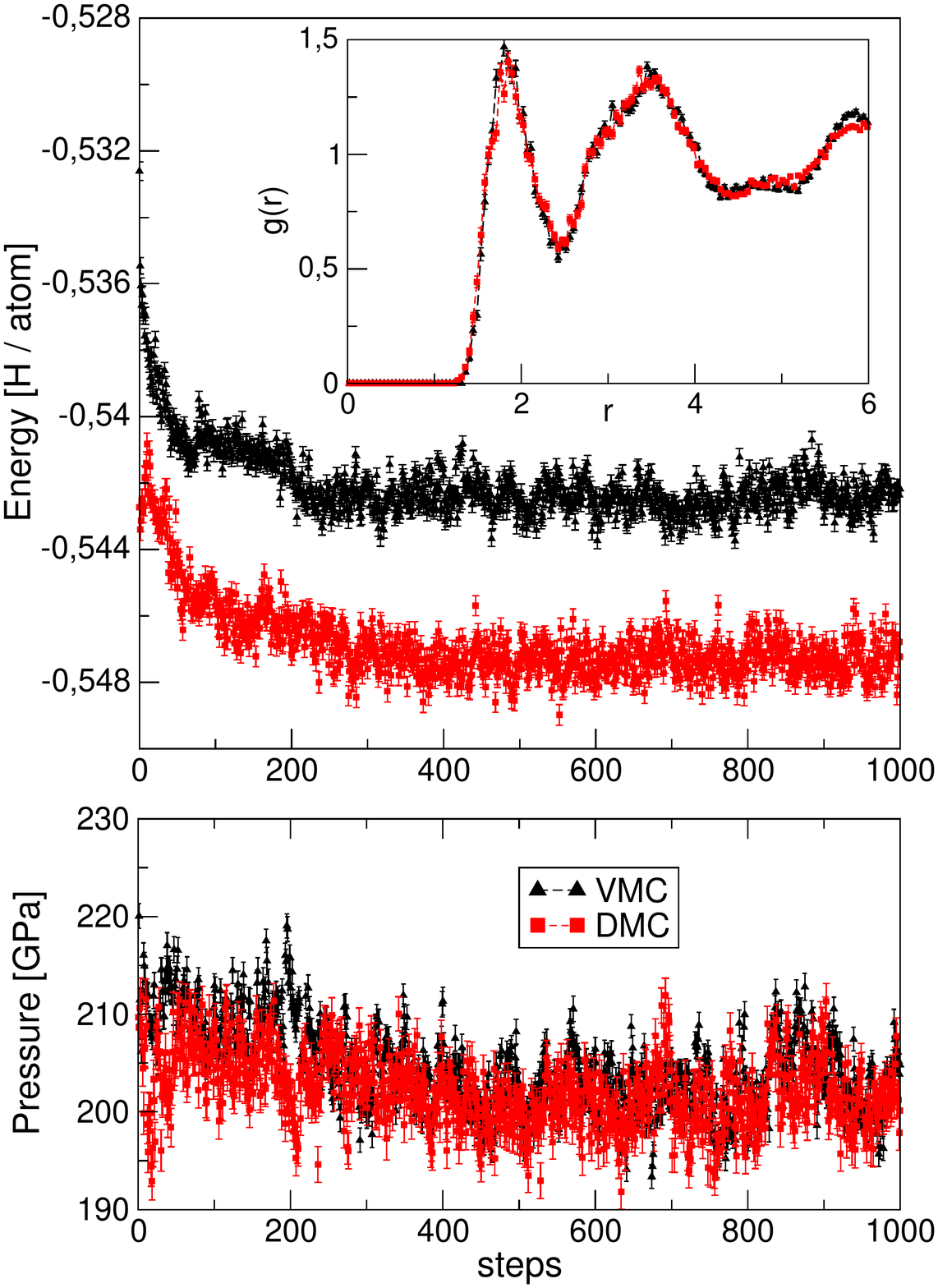}
\caption{{\bf Supplementary Figure} 2: Comparison between VMC (black-triangle up points) and DMC (square-red points) dynamics at T=600 K and density $r_s$=1.35 for a 54 proton system. In the top panel we plot the total electronic energy as a function of the (first 1000) MD steps; in the inset the radial pair distribution function. In the lower panel we plot the pressure. The average values for the pressures (evaluated after that equilibration has been achieved) are $P_{VMC}$ = 204(1) GPa and $P_{DMC}$ = 201(1) GPa. This shift in pressure at fixed density is not relevant for the present accuracy in the phase diagram. }
\label{fig:dmc}
\end{center}
\end{figure} 

\begin{figure}[h!]
\begin{center}
\includegraphics[scale=0.5]{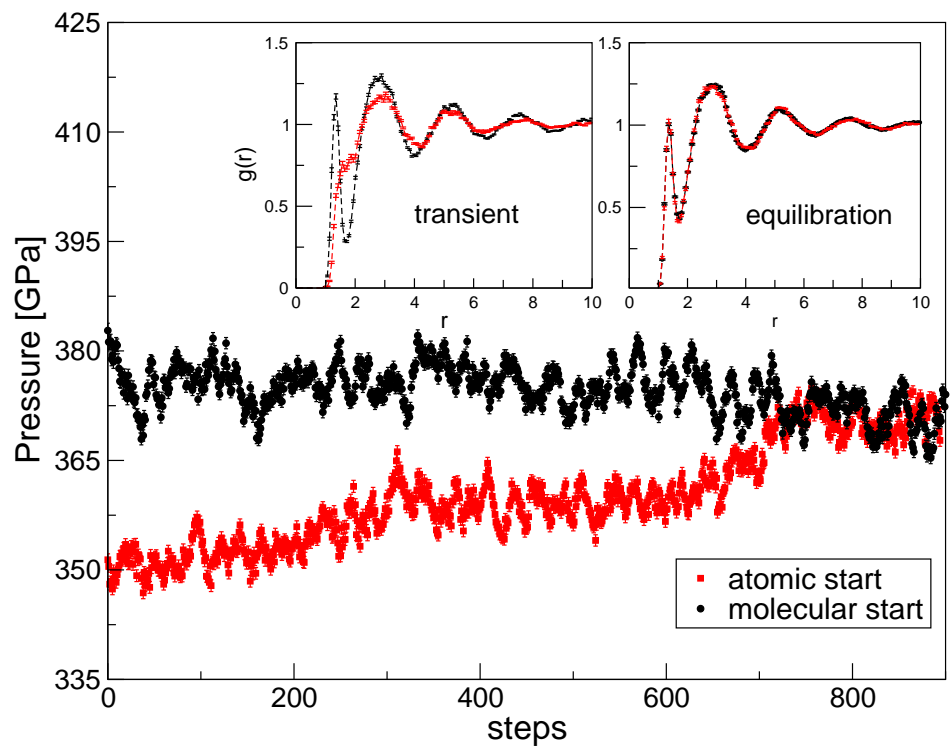}
\caption{{\bf Supplementary Figure} 3: Pressure as a function of simulation steps for two different starting configuration at T=2300 K and density $r_s$=1.32. Black points correspond to a mainly molecular initial distribution while the red ones to an atomic fluid (left inset). The two simulations thermalize halfway between the two possibilities (right inset).
The time step used in integrating the SLD is $\sim$ 0.5 fs. 256 hydrogen atoms are considered.}
\label{f:term2400}
\end{center}
\end{figure}

\begin{figure}[h!]
\begin{center}
\includegraphics[scale=0.5]{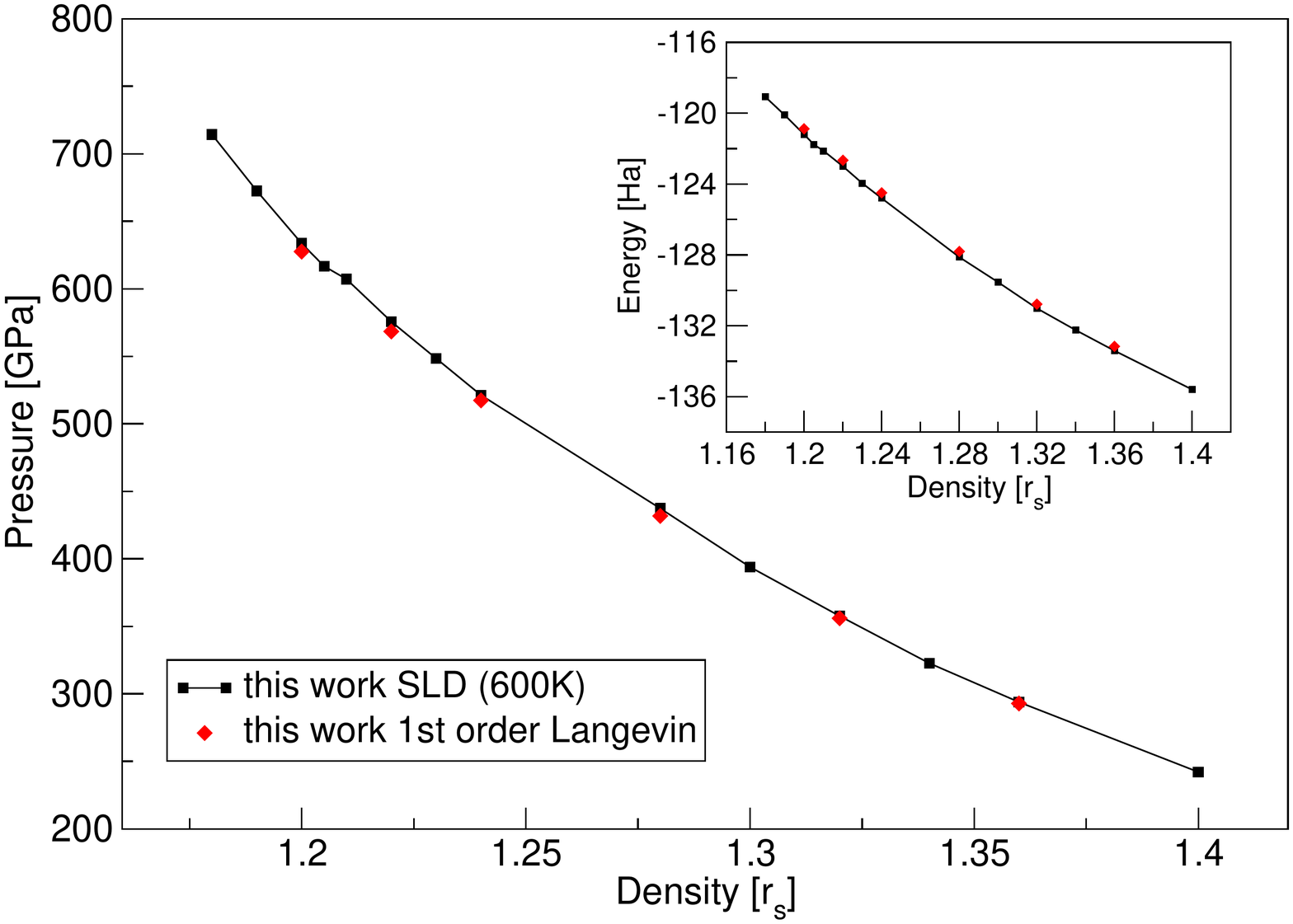}
\caption{{\bf Supplementary Figure} 4: Isotherm T=600K. Pressure as a function of density.Inset: Total energy as a function of density (the system contains 256 ions). }
\label{f:f600}
\end{center}
\end{figure}
\begin{figure}[h!]
\begin{center}
\includegraphics[scale=0.5]{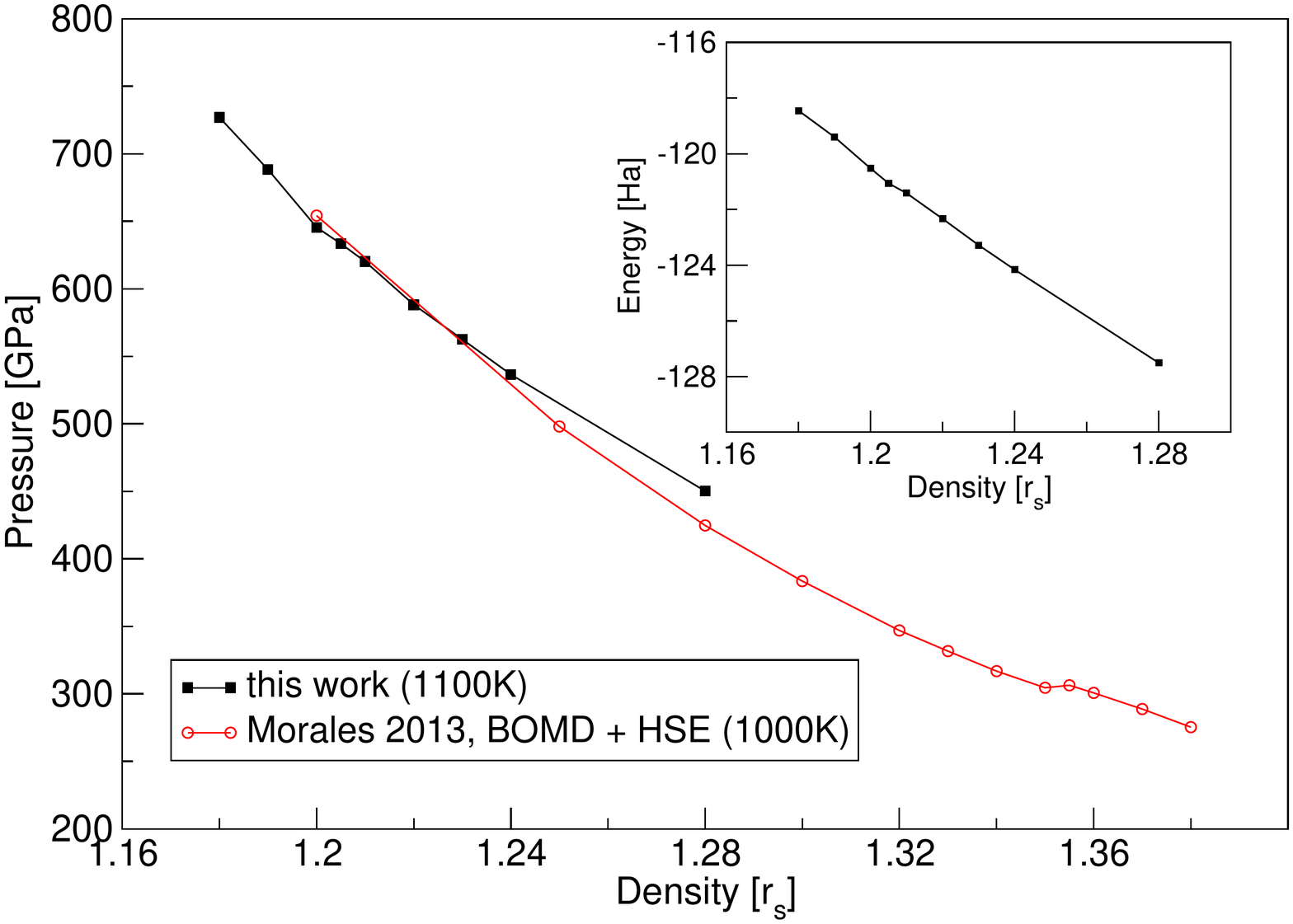}
\caption{{\bf Supplementary Figure} 5: Isotherm T=1100K. Pressure as a function of density. DFT results from Ref.\cite{mcmahon} at T=1000 K are also plotted. Inset: Total energy as a function of density (the system contains 256 ions).}
\label{f:f1200}
\end{center}
\end{figure}
\begin{figure}[h!]
\begin{center}
\includegraphics[scale=0.5]{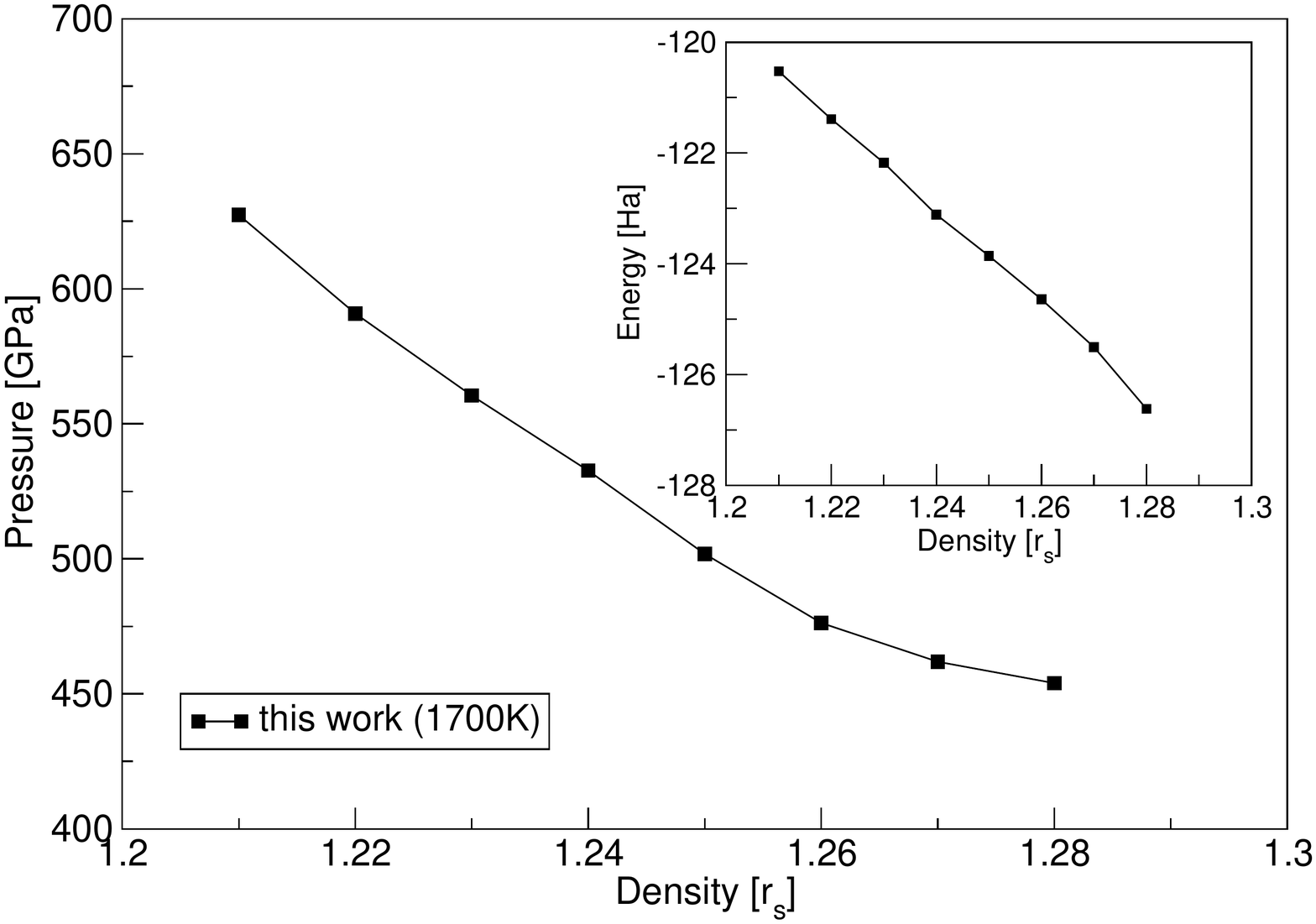}
\caption{{\bf Supplementary Figure} 6: Isotherm T=1700K. Pressure as a function of density. Inset: Total energy as a function of density (the system contains 256 ions).}
\label{f:f1800}
\end{center}
\end{figure}
\begin{figure}[h!]
\begin{center}
\includegraphics[scale=0.5]{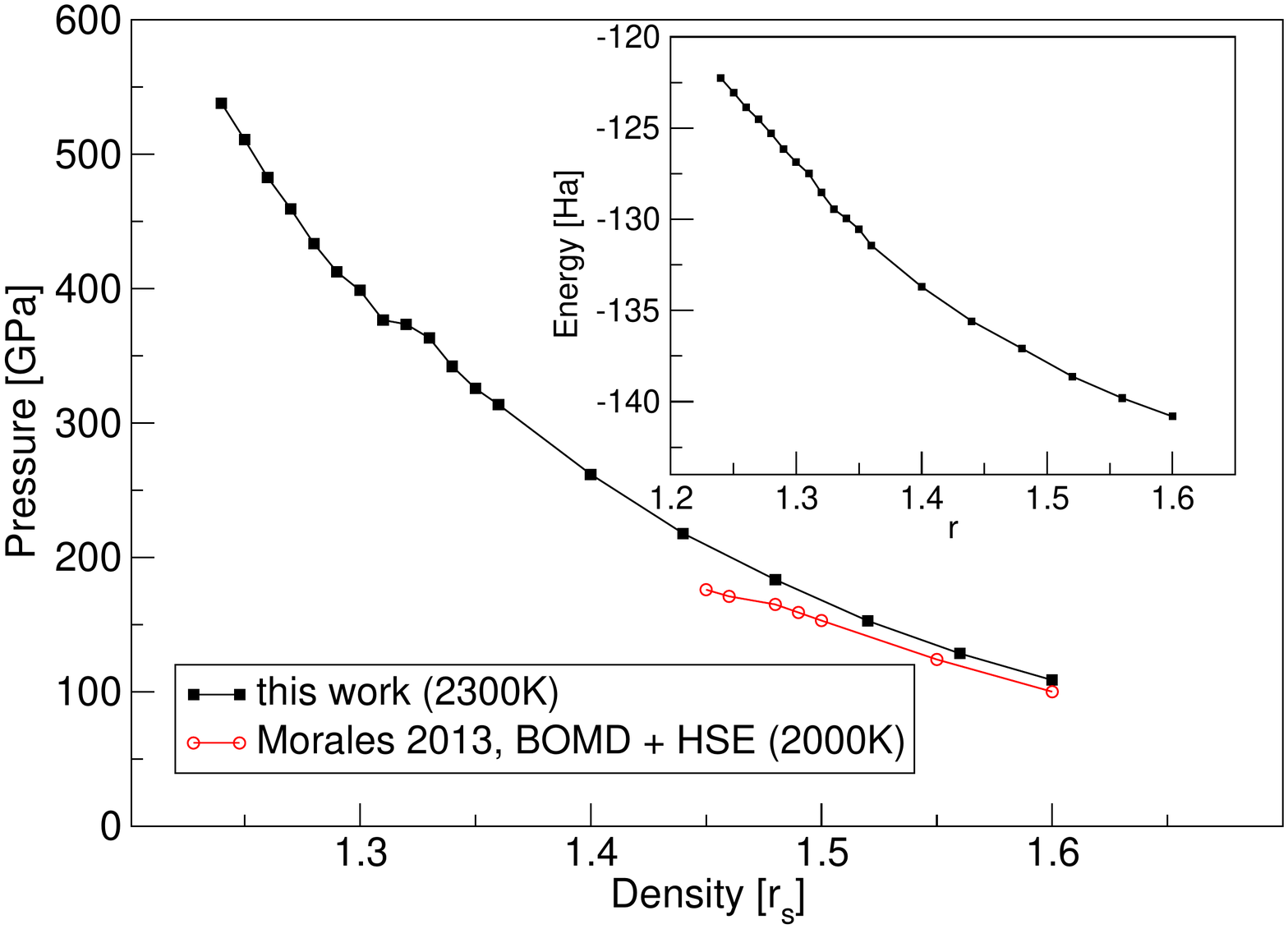}
\caption{{\bf Supplementary Figure} 7: Isotherm T=2300K. Pressure as a function of density. DFT results from Ref.\cite{mcmahon} at T=2000 K are also plotted. Inset: Total energy as a function of density (the system contains 256 ions).}
\label{f:f2400}
\end{center}
\end{figure}

\begin{figure}[h!]
\begin{center}
\includegraphics[scale=0.5]{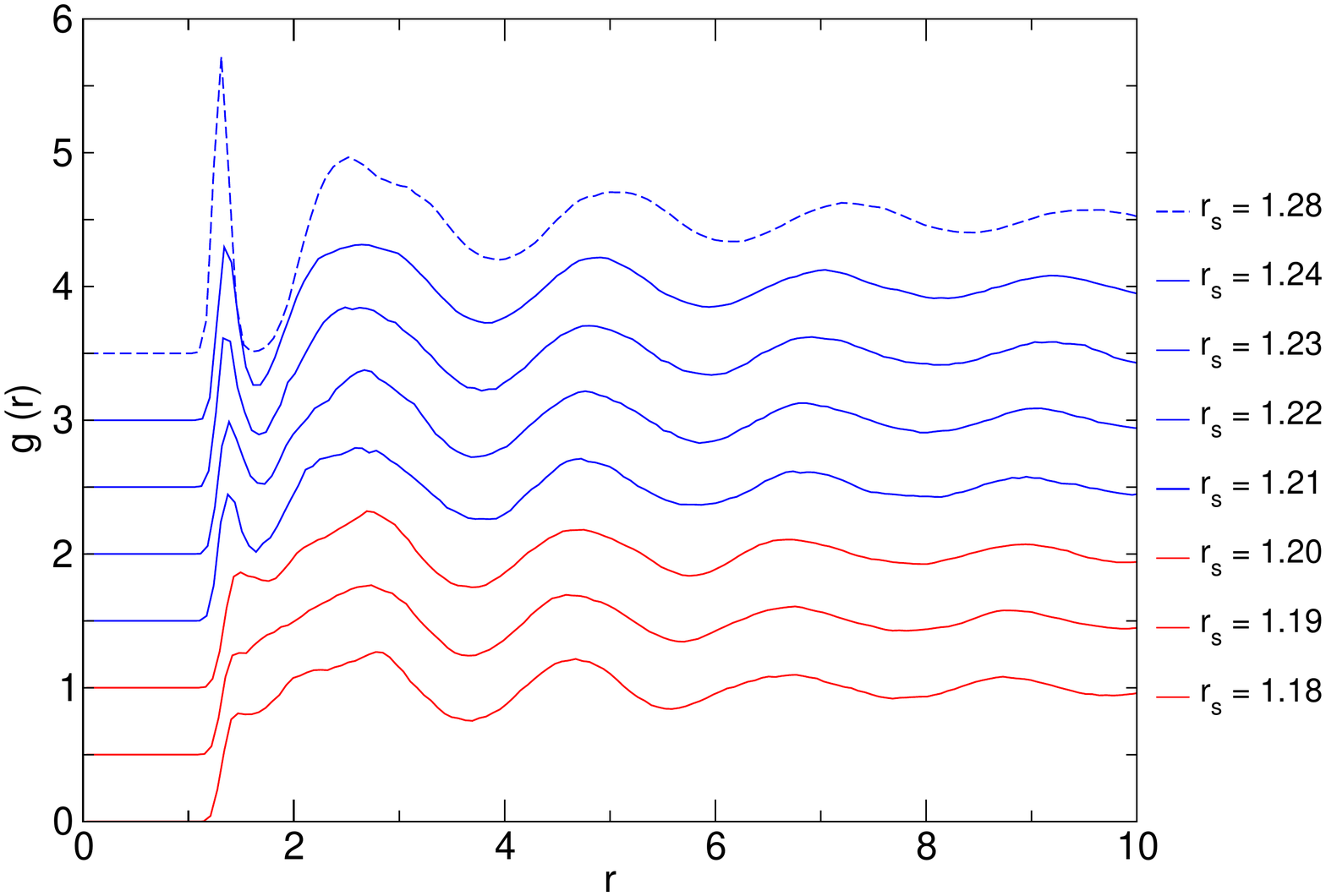}
\caption{{\bf Supplementary Figure} 8: Isotherm T=600K. Radial pair distribution function $g(r)$ for ions. The jump between the completely atomic fluid and a partially molecular one occurs between $r_s$=1.20-1.21 }
\label{f:g600}
\end{center}
\end{figure}
\begin{figure}[h!]
\begin{center}
\includegraphics[scale=0.5]{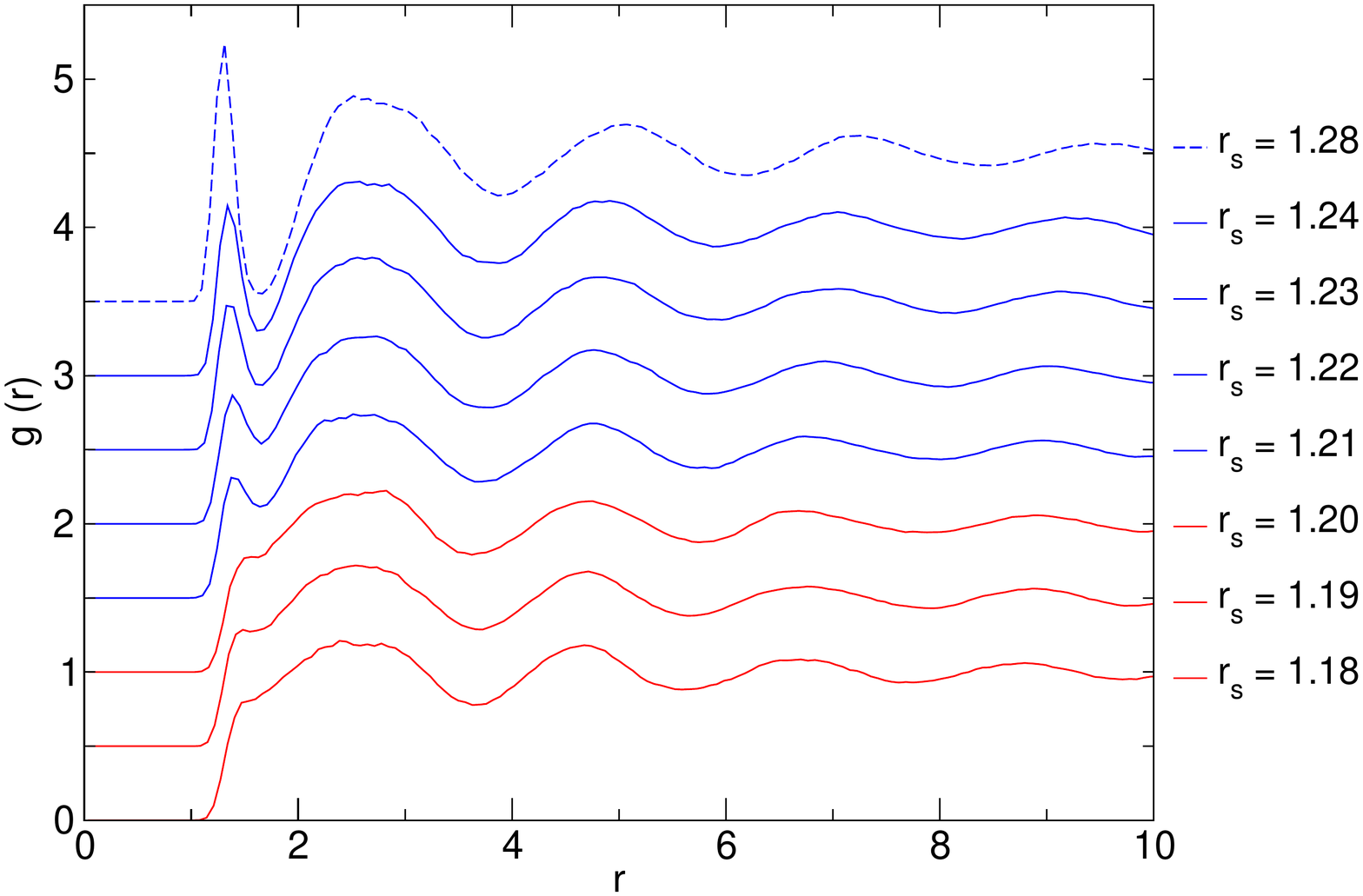}
\caption{{\bf Supplementary Figure} 9: Isotherm T=1100K. Radial pair distribution function $g(r)$ for ions. The jump between the completely atomic fluid and a partially molecular one occurs between $r_s$=1.20-1.21 }
\label{f:g1200}
\end{center}
\end{figure}
\begin{figure}[h!]
\begin{center}
\includegraphics[scale=0.5]{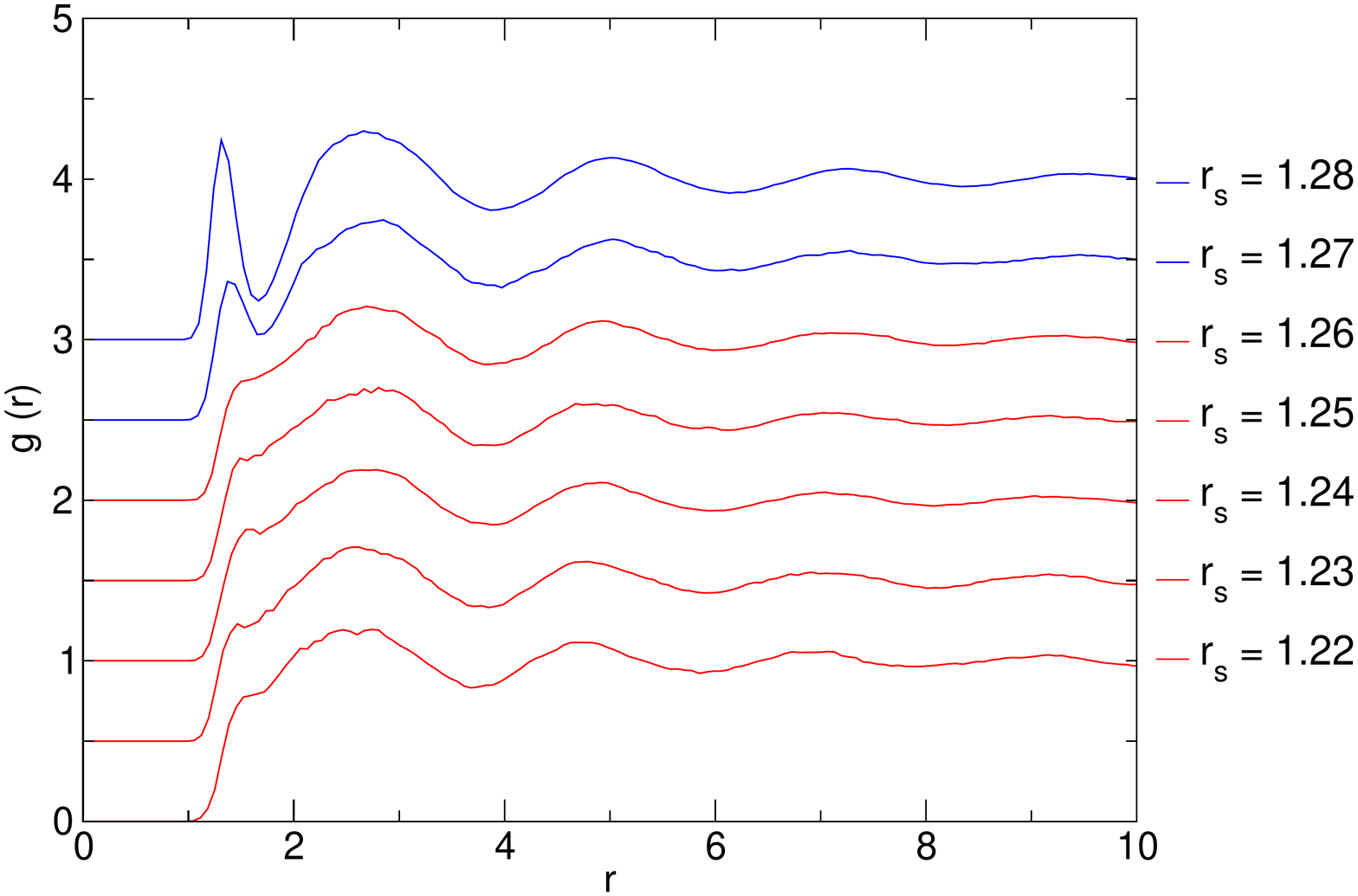}
\caption{{\bf Supplementary Figure} 10: Isotherm T=1700K. Radial pair distribution function $g(r)$ for ions. The jump between the completely atomic fluid and a partially molecular one occurs between $r_s$=1.26-1.27 }
\label{f:g1800}
\end{center}
\end{figure}
\begin{figure}[h!]
\begin{center}
\includegraphics[scale=0.5]{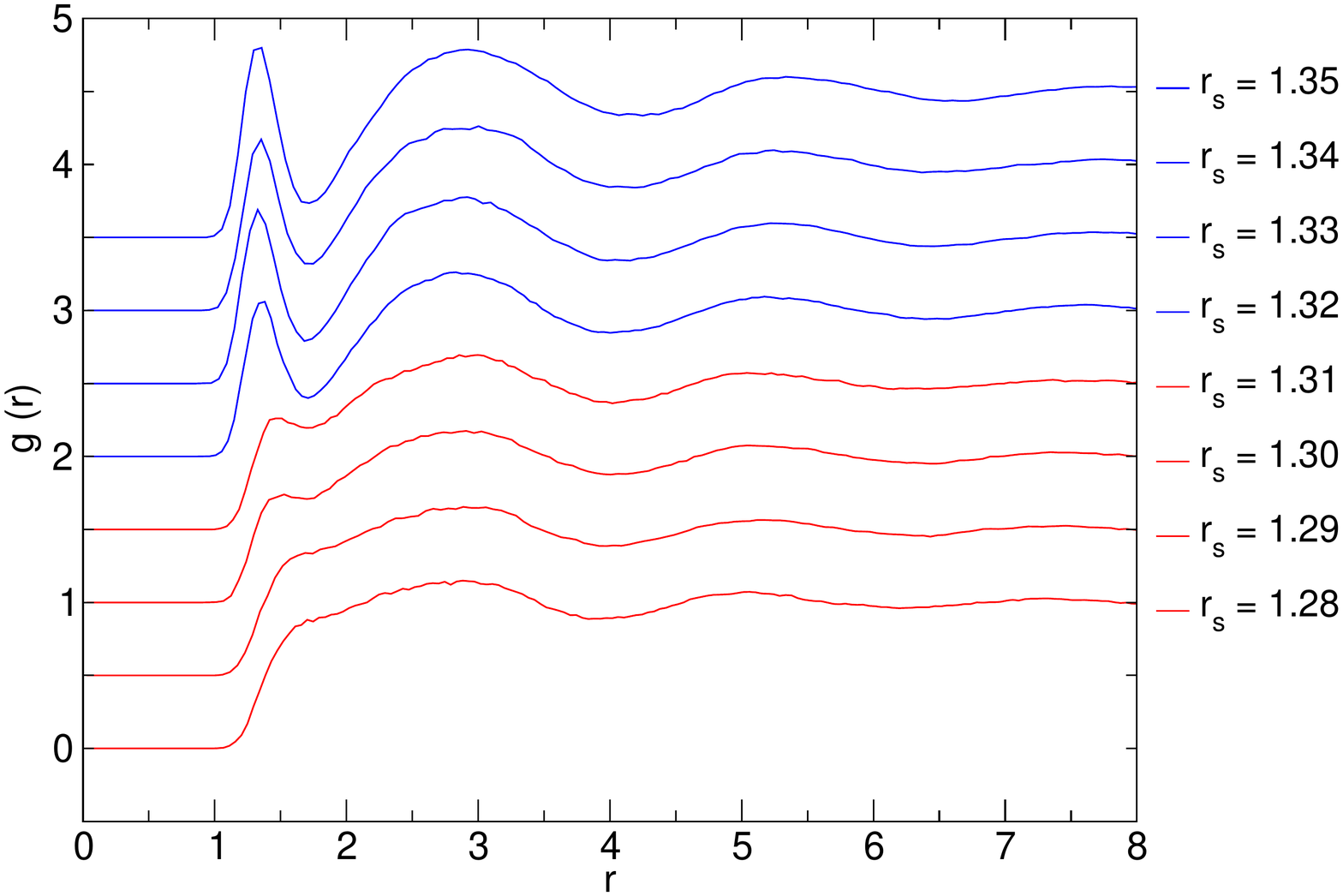}
\caption{{\bf Supplementary Figure} 11: Isotherm T=2300K. Radial pair distribution function $g(r)$ for ions. The jump between the completely atomic fluid and a partially molecular one occurs between $r_s$=1.31-1.32 }
\label{f:g2400}
\end{center}
\end{figure}

\begin{figure}[h!]
\begin{center}
\includegraphics[scale=0.5]{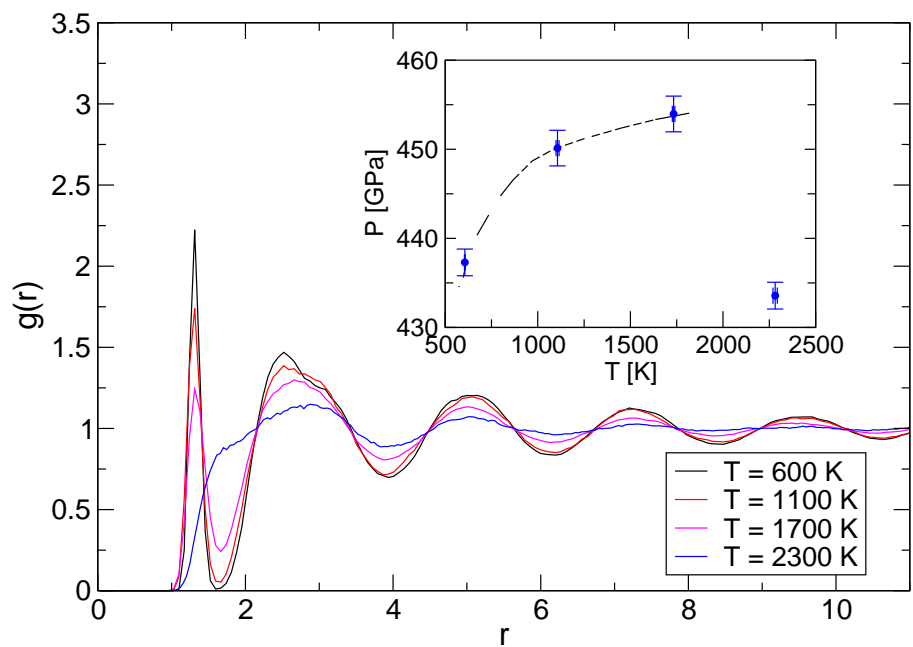}
\caption{{\bf Supplementary Figure} 12: Radial pair distribution functions for different temperatures at fixed $r_s=1.28$ density. The LLT along this isochor occurs between 1700 and 2300 K.
Inset. Pressure as a function of the temperature. The pressure increases as long as the fluid remains molecular. A drop in the pressure occurs at the dissociation.}
\label{f:conf}
\end{center}
\end{figure}

\begin{figure}[h!]
\begin{center}
\includegraphics[scale=0.5]{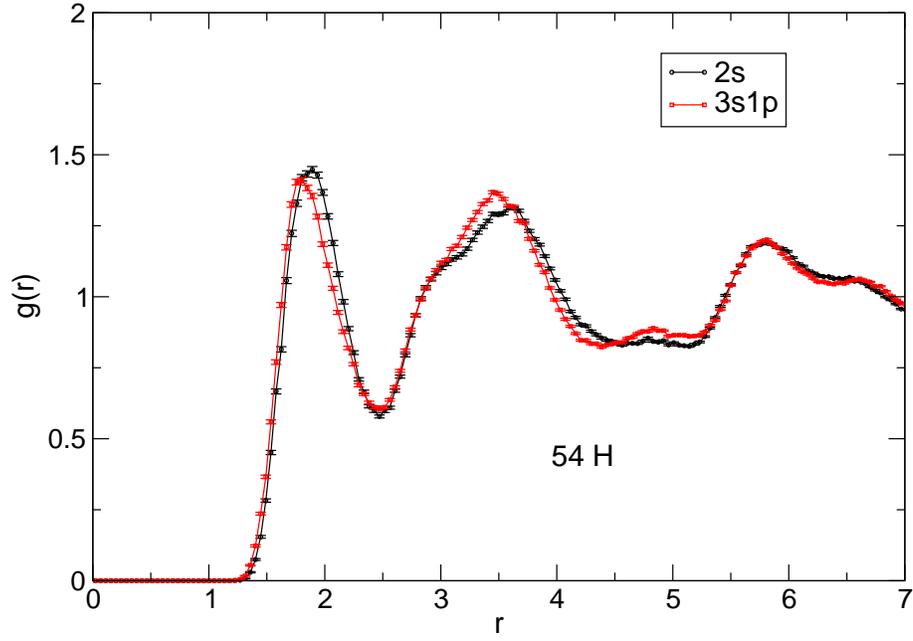}
\caption{{\bf Supplementary Figure} 13: Convergence of the ionic radial pair distribution function with respect to the basis set for a small simulation box containing 54 atoms at T=600 K and density $r_s$=1.35.
Black points correspond to a $2s$/atom localized basis set for the determinant, i.e, the one used in the result reported in the main text, while the red ones correspond
to a larger basis set $3s1p$ centered on each atom. The form of the Jastrow function is kept fixed.}
\label{f:base}
\end{center}
\end{figure} 

\begin{figure}[h!]
\begin{center}
\includegraphics[scale=0.5]{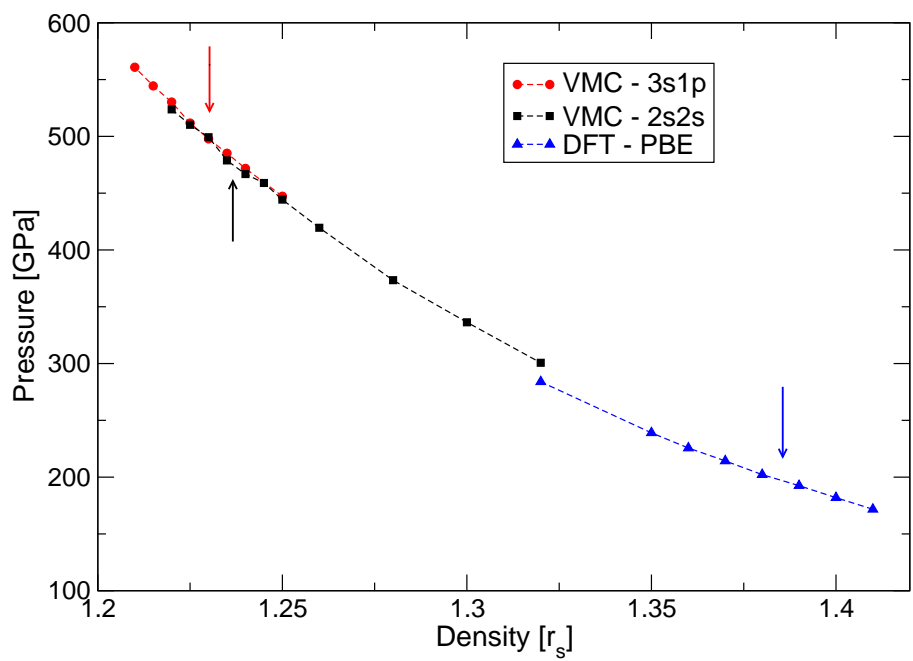}
\caption{{\bf Supplementary Figure} 14: Pressure vs density plot for a 64 atoms system at $1100K$.
Black points correspond to a $2s$/atom localized basis set for the determinant, i.e, the one used in the result reported in the main text, while the red ones correspond
to a larger basis set $3s1p$ centered on each atom. The form of the Jastrow function is kept fixed. Blue point correspond to DFT simulation with PBE functional and $\Gamma$ point. Arrows indicate the LLT (located also by looking at the radial pair distribution function). It is important to note that the LLT pressure is shifted in QMC towards higher values with respect to DFT because the LLT critical density is also shifted, the curve P vs $\rho$ being qualitatively the same.}
\label{f:basepress}
\end{center}
\end{figure} 

\begin{figure}[h!]
\begin{center}
\includegraphics[scale=0.5]{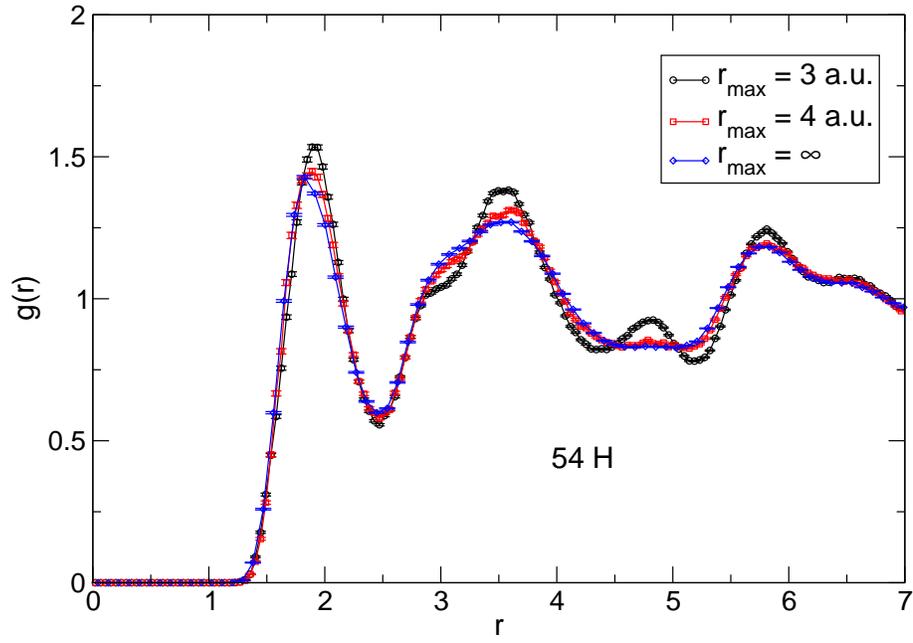}
\caption{{\bf Supplementary Figure} 15: Convergence of the ionic radial pair distribution function with respect to $r_{max}$ for a small simulation box containing 54 atoms at T=600 K and density $r_s$=1.35.}
\label{f:rmax}
\end{center}
\end{figure}

\begin{figure}[h!]
\begin{center}
\includegraphics[scale=0.5]{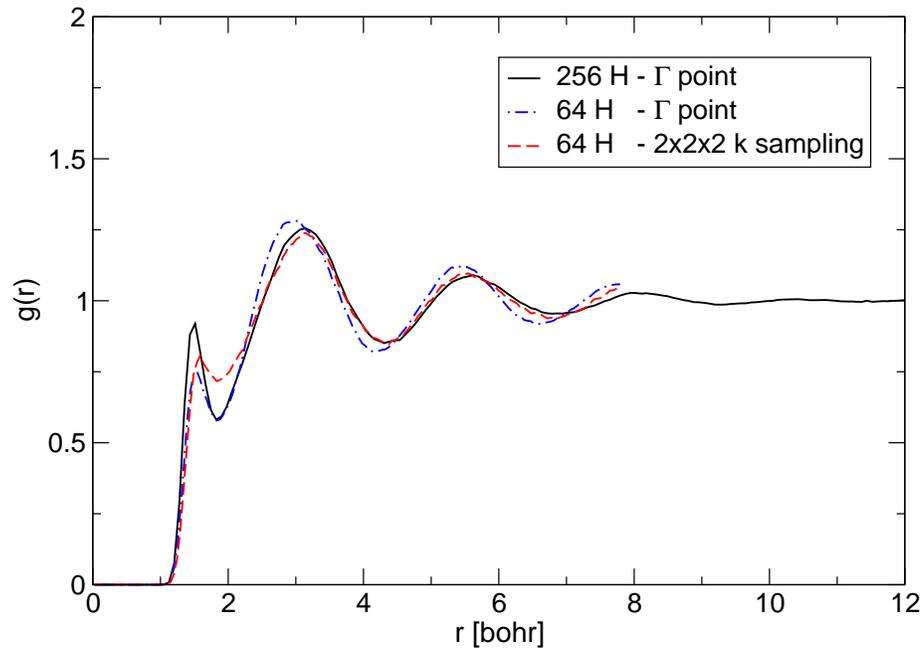}
\caption{{\bf Supplementary Figure} 16:  Convergence of the ionic radial pair distribution function at $r_s = 1.40$ and T=1100 K, with respect to the size of the supercell in DFT simulation with PBE functional. Blue line correspond to a system of 64 atom and calculation at $\Gamma$ point; the average pressure is 182(1) GPa. Red line refers to a system of 64 atom, employing a 2x2x2 sampling of the Brillouin zone; the average pressure is 196(1) GPa. Black line refers to a 256 atom simulation at $\Gamma$ point; the pressure is 206(1) GPa. The finer K-sampling seems to speed up the convergence of the pressure but does not change the nature (molecular or atomic) of the liquid.}
\label{f:dftsize}
\end{center}
\end{figure} 
\clearpage
\begin{figure}[h!]
\begin{center}
\includegraphics[scale=0.3]{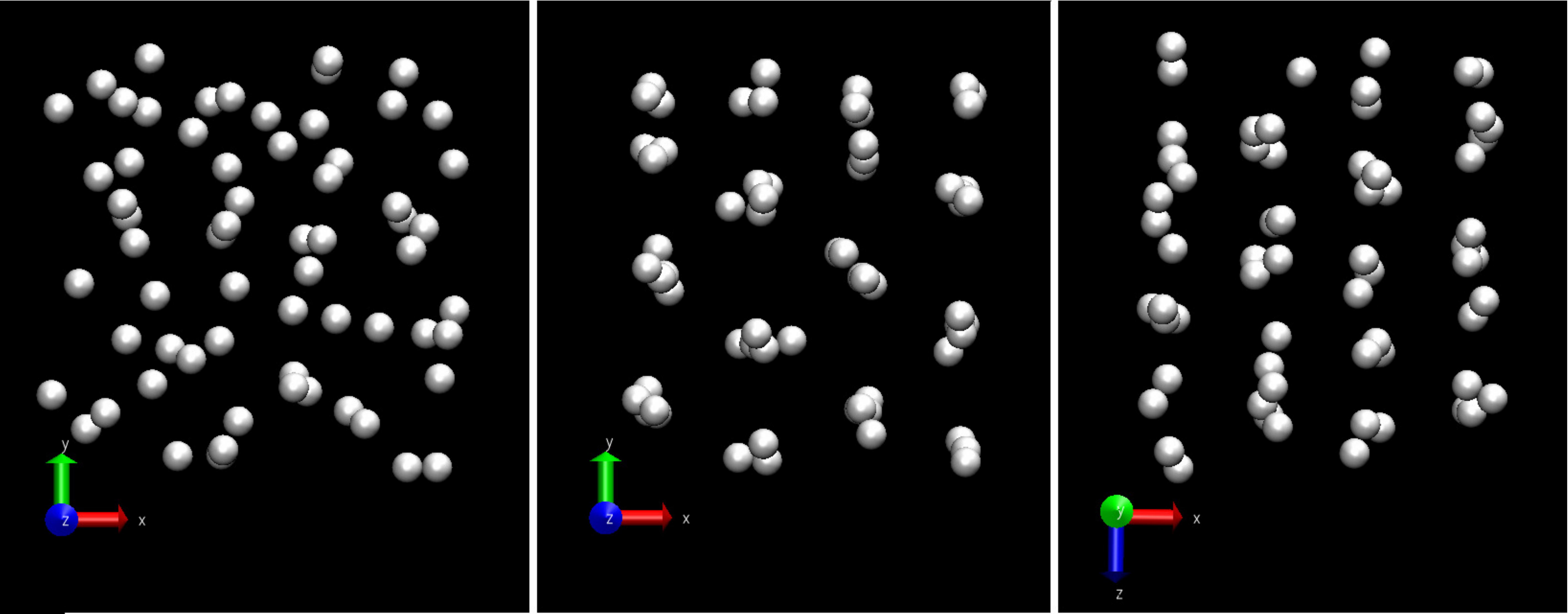}
\caption{{\bf Supplementary Figure} 17: Solidification of the liquid at 600 K and pressure 206 GPa (input density $r_s$=1.35). \emph{Left panel}: Snapshot of the initial ionic configuration. \emph{Middle panel}: snapshot of the final configuration after  $\sim 1$ ps of dynamics. \emph{Right panel}: Final ionic configuration viewed from a different angle (see axis coordinates on the bottom left of each panel).  }
\label{f:solid}
\end{center}
\end{figure}

\clearpage

\begin{table}[h!]
\centering 
\begin{tabular}{c c c c} 
\hline\hline 
$r_{max}$ &$|$ \# parameters & $|$  Energy error \\ [0.5ex] 
\hline 
0 & 1542 &   0.022(6) \\
2 & 1813 &   0.0044(11)\\
3 & 3003 &   0.001(1) \\
4 & 5108 &   0.004(1)\\
5 & 8443 &   -  \\ [1ex] 
\hline 
\end{tabular}
\label{table:nonlin} 
\caption{{\bf Supplementary Table} 1:  Optimized energy for a 256 atom equilibrated configuration as a function of $r_{max}$. Energies are given as a difference with respect to the $r_{max}=5$ reference value $E(r_{max}=5)=-128.236(4) H$. It is noteworthy that even the $r_{max}=0$ setting improve upon the J+DFT total energy -128.157(4), in which molecular orbitals are not optimized in the presence of the Jastrow factor. }
\end{table}

\begin{table}[h!]
\centering 
\begin{tabular}{c c c c} 
\hline\hline 
k point mesh & $|$ $r_s=1.40$ (201 GPa) & $|$ $r_s=1.35$ (259 GPa) \\ [0.5ex] 
\hline 
$\Gamma$ & -0.5267 &  -0.5168 \\
2$\times$2$\times$2 & -0.5278 &  -0.5181\\
3$\times$3$\times$3 & -0.5277 &  -0.5180 \\
4$\times$4$\times$4 & -0.5277 &  -0.5180\\
\hline
$|\Delta|$ & 0.0010 &   0.0012  \\ [1ex] 
$\Delta$/E  & 0.19\% &   0.23\%  \\ [1ex]
\hline 
\end{tabular}
\label{table:ksamp} 
\caption{{\bf Supplementary Table} 2: Convergence of the DFT total energy (H/atom) as a function of the number of $k-$ points for two static 256 proton configurations obtained from two different equilibrated DFT-MD simulations with PBE xc functional at different densities and T=1100 K. 
The first configuration is representative of the molecular liquid ($r_s=1.40$ (201 GPa)) while the other of an atomic liquid ($r_s=1.35$ (259 GPa)) and they are both close to the liquid-liquid transition at T=1100 K.
We define $\Delta$ as the energy difference between the $\Gamma$ point calculation and the 4$\times$4$\times$4 fully converged one. Notice that $\Delta$ is negligible with respect to the total energy of the configurations  and remains much smaller than the energy difference $ \delta E=-0.5277+0.5180=0.0097$ H/atom between the molecular and atomic configurations. Moreover, since $\Delta$ is comparable in the two phases the bias in the calculation of the energy differences in the two phases is only 60 K/atom $\sim 1/50 \delta E$.
 Therefore  the $\Gamma$ point approximation on this large system size is justified as the use of a better $k-$point sampling will  likely shift the transition pressure by  at most  few GPa's which are not relevant for the proposed phase diagram.
}
\end{table}

\clearpage

\section{Supplementary Note 1: Details of the sampling scheme}

{\bf Second order Langevin equation.} As mentioned in the main text, we perform ab-initio molecular dynamics (MD) simulations via variational quantum Monte Carlo (VMC).
We employ \emph{TurboRVB} QMC package (http://people.sissa.it/\textasciitilde sorella/web/).
A second order Langevin dynamics (SLD) is used in the sampling of the ionic configurations, within ground state Born-Oppenheimer approach.
Ionic forces are computed with finite and small variance,
which allows the simulation of a large number of atoms. Moreover the statistical noise, corresponding to the
forces, is used to drive the dynamics at finite temperature by means of an appropriate generalized Langevin dynamics~\cite{attaccalite}.
A similar approach has been proposed  in Ref.\cite{par1} and Ref.\cite{par2} where a SLD algorithm has been devised also at the DFT level.
In this work we adopt a different numerical integration scheme for the SLD which allows us to use
large time steps, even in presence of  large friction matrices.
The integration of the SLD follows the same rational of the original paper~\cite{attaccalite}, 
for solving a set of differential equations:
\begin{eqnarray} 
 \dot {\bm v }  &=&  -   \gamma
  \bm v   +   \bm f(\bm R)  +\bm \eta  (t) \label{first} \\
 \dot{  \bm R }   &=&  \bm v  \label{second}   
\end{eqnarray}
where $\bm R, \bm v, \bm f$ are respectively the positions, the velocities and the forces, and the noise $\eta$ is determined by the fluctuation-dissipation theorem, namely its instantaneous 
correlation $\alpha$ is given by:
\begin{equation}
\alpha(\bm R) = 2 T \gamma(\bm R)
\end{equation}
where $T$ is the temperature and both $\gamma(\bm R)$ and $\alpha(\bm R)$ are $3M$ dimensional square matrices, 
$M$ being the number of ions, whereas $R$ indicates here the $3-M$ dimensional vector made by the 
positions of all the atoms.
Since one of the two matrices is arbitrary, we choose $\alpha(\bm R)$ in the following form:
\begin{equation}
\alpha = \alpha_0 I + \Delta_0 \alpha_{QMC}(\bm R)
\end{equation}
where $\alpha_{QMC}(\bm R)$ 
defines the correlation of the forces in QMC, $I$ is the identity matrix and $\alpha_0$ is a constant that 
should be optimized to minimize  the autocorrelation time and therefore the efficiency of the  sampling.
At variance of the original paper, here, in order to be more accurate, 
we do not use the simplest approximation 
to solve the Eq.(\ref{second}), because the velocity, when the friction matrix $\gamma$ is large 
can have strong variations in the discrete time  integration step $\Delta$.
Indeed, we now  assume only that in the interval $ t_n-\Delta/2  < t < t_n+\Delta/2$,
the positions
$\bm R$ are changing a little and, within a good approximation,
 we can neglect the $R$ dependence in the RHS of Eq.(\ref{first}).
Moreover the velocities
$\bm v_n$ are computed at half-integer  times $t_n-\Delta/2$, whereas
coordinates $\bm R_n$ are assumed to be defined at integer times
$\bm R_n= \bm R(t_n)$.
Then the solution can be given in a closed form:
\begin{eqnarray}
 \label{e:disc_in}
\bm v_{n+1}  &=&e^{ -  \gamma \Delta  }  \bm v_{n} +
\bar \Gamma  ( \bm f(\bm R_n) +   \bm { \tilde  \eta} )   \\
\bm R_{n+1} & =& \bm R_n+   e^{ -  \gamma \Delta/2   } \Gamma \bm v_n  
+ \bar \Theta ( \bm f (\bm R_n) + \bm   { \tilde { \tilde \eta} }  )  \label{eqdyn2}   \\
\bar \Gamma &=&  \gamma^{-1} ( 1 - e^{ - \gamma \Delta } )  \\ 
{\bm { \tilde \eta} } &=& {  \gamma  \over  2  \sinh( \Delta/2  \gamma) } 
\int\limits_{t_n-\Delta/2}^{t_n+\Delta/2} dt e^{ \bar \gamma (t-t_n)} \bm \eta (t)     \\ 
 {\bm { \tilde { \tilde  \eta}} } &=&  \bar \Theta^{-1}  
\int\limits_{t_n}^{t_{n+1}} dt  \int\limits_{t_n-\Delta/2}^{t} d\tau e^{  \gamma (\tau-t)} \bm \eta (\tau)     \\ 
\bar \Theta &= &  \gamma ^{-1} ( \Delta - e^{- \Delta/2 \gamma} \bar 
\Gamma )
\end{eqnarray}
 By using  that
$\left[  \alpha, \gamma\right]=0$ and a little algebra,  
  the correlator defining the discrete (time integrated)
noise can be computed and gives:
\begin{eqnarray} \label{totalnoise}
< \tilde \eta_i \tilde \eta_j > & = &{  \bar \alpha^{1,1}  } \\
 &=&   \alpha   \gamma  {  \sinh ( \Delta  \gamma) \over
 4 \sinh( \Delta  \gamma/2)^2 }  \nonumber   \\
< { \tilde { \tilde \eta}}_i { \tilde { \tilde \eta}}_j > & = &
{\bar \alpha^{2,2} }  \\ 
&=& 
 \alpha \bar \Theta^{-2}  \left[ { \Delta \over  \gamma^2 } \right. \nonumber  \\ 
&+& \left.  {1 \over 2  \gamma^3} \left( -2 + e^{ -\Delta  \gamma } 
+2 e^{ - 2  \gamma \Delta } - e^{ - 3  \gamma \Delta } \right) \right] 
\nonumber \\
< { \tilde { \tilde \eta}}_i  { \tilde \eta}_j > & = & {  \bar \alpha^{2,1}  }  = {  \bar \alpha^{1,2}} \\
&=&  {  \alpha \bar \Theta^{-1} \over 
4 \bar \gamma \sinh( { \Delta \bar \gamma \over 2 }) } \nonumber \\ 
&&  \left( 2 e^{\bar \gamma \Delta \over 2 } -2 - e^{-\bar \gamma \Delta} 
+ e^{ - 2 \bar \gamma \Delta } \right) \nonumber 
\nonumber \\
\label{e:disc_end}
\end{eqnarray}
As it is seen, the equations determining  the noise correlations 
 are now more complicated, as they involve 
a $2\times 2$ block matrix $\bar \alpha^{i,j}$, where each block is a $3M \times 3M$ submatrix.
Apart for this, the generalization of the noise correction to this case is straightforward, 
as to each of the four submatrices we have to subtract the $3M\times 3M$ QMC correlation of the 
forces $\alpha_{QMC}$, namely $\bar \alpha^{ij}_{ext} =  \bar \alpha^{ij} - \alpha_{QMC}$ is 
the true external noise we have to add to the system, to take into account that QMC forces 
contains already a correlated noise.
It can be shown, by a simple numerical calculation, 
 that the resulting matrix $\bar \alpha_{ext}$  is  indeed positive definite provided 
$ \Delta_0  > { 4 \over 3} \Delta $, so that $\bar \alpha_{ext}$ is a well defined 
correlation for an external noise.

We test the convergence of this improved  numerical scheme against the standard Euler discretization scheme (which is affected by the constraint $\Delta \gamma < 1$)
 on an analytically solvable classical toy model as shown in Supplementary Figure (\ref{f:toy}). In this example forces are computed analytically, i.e. $\alpha_{QMC} = 0$, nevertheless
the gain obtained by the discretization scheme (\ref{e:disc_in}-\ref{e:disc_end}) is clear.

{\bf First order Langevin dynamics.} We perform also simulations using the standard first order Langevin dynamics to further validate our results.
In this case the updates of ionic coordinates $\bm R$ are given by
\begin{eqnarray}
 \dot {\bm R}  & = \bm f (\bm R)  + \bm \eta 
\end{eqnarray}
with $ \langle  \eta_i(t)  \eta_j(t^\prime)  \rangle  =\delta(t-t^\prime) \delta_{i,j} ~ { 2  T } $.
Simulations within this framework are much more expensive since the forces must be evaluatued with a statistical noise $\eta_f \Delta$ much smaller 
than $\sqrt{2 T \Delta}$. 
Indeed the Monte Carlo noise in the forces may give a significant bias to the effective temperature as the noise correction technique is not available in this scheme.
Moreover this first order Langevin dynamics is found to give samples $\bm R(t)$ much more correlated than the SLD defined above so longer 
runs are required to sample correctly the whole configuration space.
We perform several simulations along the T=600 K isotherm as shown in Supplementary Figure \ref{f:f600}. These test simulations are consistent with the SLD results, but require  at least an order of magnitude more computational resources.
This also gives us the insight that the time-step discretization error is under control since two different integration schemes (1st and 2nd order) provide the same outcome.

{\bf Setting up the MD for realistic simulations.} For the realistic $N-$hydrogen systems we integrate this SLD (with this novel discretization scheme) using a time step of $\sim 0.5-1.0$ fs, the larger the temperature the smaller the time step used.
For each step of molecular dynamics we employ six iterations of wave function optimization using 
the so called linear method~\cite{rocca,umrigar}, that, we have checked, allows us to satisfy rather well 
the Born-Oppenheimer constraint of minimum energy at fixed ionic positions. 
 Simulations last long enough until thermalization is established. A typical run at fixed density and temperature is  about $2$ ps long.
For each of the four isotherm, i.e. 600, 1100, 1700, 2300 K, we perform several simulations varying the density, with a mesh increment of 0.01 for $r_s$.
$r_s$ is the Wigner-Seitz radius defined by $V/N= 4/3 \pi (r_s a_0)^3$
where $V$ is the volume, $N$ the number of ions, and $a_0$
is the Bohr radius.
As described in the main text we perform careful checks on the equilibration in order to avoid hysteresis effects along each isotherm.
Here (see Supplementary Figure \ref{f:term2400}) we report also the thermalization steps in a SLD at 2300 K and $r_s$=1.32, i.e. on the molecular edge of the LLT at this temperature.
As it is seen, a simulation in which the fluid remains completely dissociated would underestimate the pressure at this density.
Thus the lack of thermalization can easily shift the LLT by several tens of GPa's.

{\bf Calculation of pressure. } The calculation of pressure is done by an expression more accurate than the standard use of the virial theorem. The energy change due to an infinitesimal volume change  $L\to L+dL$ in a box of volume $L^3$ 
can be exploited more conveniently 
by first scaling all electronic and ionic coordinates by $L$ and mapping them onto
  a cubic box of unit size.
After this  scaling, a very simple parametric dependence on $L$ 
appears in the scaled Hamiltonian  and the wave function.     
The conventional virial expression turns out in a simple way by considering only the Hamiltonian 
dependence in the standard expression for the derivative of the VMC energy. 
This is an approximation at the variational level, because 
one cannot neglect the ''weak'' dependence on $L$ of the wave function in the important part of the Jastrow 
 necessary to satisfy the cusp conditions.\cite{marchi} 
Therefore  we have considered here the correct expression of the pressure, that at the variational 
level corresponds to the exact energy derivative with respect to the above mentioned infinitesimal volume 
change.
\clearpage

\section{Supplementary Note 2: Molecular dynamics with Fixed Node Diffusion Monte Carlo}
In this section we describe a simple way to extend the present molecular dynamics
with an approximation, the so called Fixed Nodes Diffusion Monte Carlo (FNDMC), which is considered the state of the art within QMC techniques for fermions, and 
that we will briefly describe below.
Given a wave function $\Psi_T$ it is possible to filter out its ground state 
component with a projection technique based on the imaginary time propagation 
$ | \Psi \rangle \to \exp(- H \tau ) |\Psi_T \rangle $, for large imaginary time 
$\tau$. Unfortunately, for fermions this projection 
is unstable for large imaginary time due to the unfamous ''sign problem'' 
instability.
Therefore, a very successful workaround has been proposed long time 
ago\cite{alder}, namely the above imaginary time  
propagation is restricted with the condition that the nodes of the 
wave function do not change in time. With this restriction the simulation turns  
again very stable but approximate. It remains a  
 variational method and in practice turns out to be 
very accurate even for strongly 
correlated systems, such as the Jellium gas at very low density.\cite{jellium}
The better are the nodes of the initial wave function and the more accurate the 
approximation is.

In Supplementary Figure(\ref{fig:dmc}), we see that the DMC considerably improves the total energy but does not 
appreciably change the quantities that are important for this work.
Namely the pressure is essentially unchanged 
(Supplementary Figure \ref{fig:dmc} lower panel),
and the $g(r)$ remains quantitatively the same in all range of distances considered (see inset Supplementary Figure \ref{fig:dmc}).
On the other hand the DMC algorithm is about an order of magnitude slower 
than the VMC, and larger number of atoms are not possible with the  
available computational resources.

\clearpage

\section{Supplementary Note 3: Complete outcomes of the simulations}

We report here in details the outcomes of the simulations. 
An extensive range of densities and pressures is scanned along the T=600 and T=2300 isotherms. The LLT extimated at this two ends helps to reduce the number of simulations required to identify the LLT in the other two inner isotherms.
In Figs. (\ref{f:f600},\ref{f:f1200},\ref{f:f1800},\ref{f:f2400}) we report the pressure and internal energy vs density plots. For T=2300 K and T=1100 K points obtained by DFT simulations as in Ref.\cite{mcmahon} are also plotted. This DFT points refer to BOMD, i.e. with classical ions as in our work, with HSE DF. Although the temperatures are not exactly matching, the two curves  display a similar P vs $\rho$ behavior,
 but in our work the LLT occurs at higher densities.
Since the discontinuity in the pressure appears to be rather small, it is extremely important 
 to identify the LLT by  looking also at the radial pair distribution function $g(r)$ for the ions.
In Supplementary Figure (\ref{f:g600},\ref{f:g1200},\ref{f:g1800},\ref{f:g2400}) we report the $g(r)$'s for densities near the LLT. A clear jump is visibile and is always connected with the discontinuity in the pressure. The $g(r)$ are shifted each by 0.5 for the sake of clarity. Red lines correspond to atomic fluids, i.e at larger pressures than the LLT, while blue ones correspond to molecular -or partially molecular- liquids.
Finally we report also a pressure vs temperature plot (see. Supplementary Figure \ref{f:conf}) at fixed density, i.e. at $r_s=1.28$. Although the mesh in temperature is quite poor a discontinuity in the pressure is fairly evident supporting the first order nature of the LLT as mentioned in the main text.
All the results here reported refer to a cubic simulation box containing 256 atoms.

\section{Supplementary Note 4: Justification of the ground-state Born-Oppenheimer approach}
As discussed in the main text we perform the dynamics adopting the ground state BO approximation, namely at each ionic step the wave function is reoptimized and forces are evaluated at the electronic ground state.
Neglecting electronic entropy is a very good approximation in this range of temperatures (600-2300 K). Indeed we have checked that, at DFT level and at the highest temperature here considered (2300K), the electronic entropy contribution $-TS$ to the total free energy $F = U -TS$ is negligible in comparison not only  with the total energy, i.e. $|TS| / |U| < 0.04\%$ (in the relevant pressure range 200-400 GPa), but also with respect to the tiny internal energy variation at the transition $\sim 4 \cdot 10^{-3}$ H/atom.
This situation changes dramatically at higher temperatures.
For instance at 23000 K, (i.e. a temperature 10 times higher than the maximum T investigated in the main text) the above  $|TS|/|U|$ ratio is already $~ 4\%$ and electronic entropy effects are relevant.
DFT simulations were performed using the QuantumEspresso code\cite{qe}.

\section{Supplementary Note 5: Solidification of the fluid at low pressures}

To test the validity of this framework we perform some simulations at lower pressures where the existence of a solid phase is experimentally well established.
We experience a spontanous solidification along the isotherm $T=600$ K for pressures of about 206 Gpa, i.e, where the liquid should meet the re-entrant melting line of the solid.
Indeed a layered structure is rapidly formed within our 64 atom cubic simulation cell (see Supplementary Figure \ref{f:solid}) starting from a molecular liquid.
Altough the mixed molecular-atomic behaviour of its bond pattern is qualitatively similar to the solid phase IV,  a precise characterization of this structure would require several simulations varying either shape of the simulation cell and the number of atoms.
For the time being we can only acknowledge this result as a positive test regarding the good degree of ergodicity that characterizes this molecular dynamics.

\newpage 
\section{Supplementary References}

\end{document}